\documentclass[twocolumn]{article}
\usepackage[top=0.8in, bottom=0.8in, left=0.75in, right=0.75in]{geometry}
\usepackage{amsmath, amsthm, amssymb, amsfonts}
\usepackage{appendix}

\usepackage{etoolbox}

\AtBeginEnvironment{equation}{\small}
\AtBeginEnvironment{align}{\small}
\AtBeginEnvironment{gather}{\small}

\usepackage[english]{babel}
\usepackage{float}
\usepackage{cite}
\usepackage{graphicx}
\usepackage{algorithm, algpseudocode}
\usepackage{soul}
\usepackage[acronym, toc,numberedsection = nolabel, section = part]{glossaries}
\usepackage[binary-units,per-mode=symbol,detect-all=true]{siunitx}
\def\defaultDigitGrouping{4}
\makenoidxglossaries
\newacronym{alap}{ALAP}{as-late-as-possible}
\newacronym{asap}{ASAP}{as-soon-as-possible}
\newacronym{alu}{ALU}{arithmetic-logic unit}
\newacronym{asm}{ASM}{approximate string matching}
\newacronym{asic}{ASIC}{application specific integrated circuit}
\newacronym[\glsshortpluralkey={bp}]{bp}{bp}{base pair}
\newacronym{bpm}{BPM}{bit-parallel Myers's}
\newacronym[longplural={block random access memories}]{bram}{BRAM}{block random access memory}
\newacronym{cgra}{CGRA}{coarse-grain reconfigurable array}
\newacronym{cnf}{CNF}{conjunctive normal form}
\newacronym{cil}{CIL}{compute-intensive loop}
\newacronym{dfg}{DFG}{data flow graph}
\newacronym{dma}{DMA}{direct memory access}
\newacronym{dp}{DP}{dynamic programming}
\newacronym{dram}{DRAM}{dynamic random-access memory}
\newacronym{dsp}{DSP}{digital signal processor}
\newacronym{edp}{EDP}{energy-delay product}
\newacronym{fpga}{FPGA}{field programmable gate array}
\newacronym{fifo}{FIFO}{first-in, first-out}
\newacronym{ff}{FF}{flip-flop}
\newacronym{gpgpu}{GPGPU}{general-purpose computing on graphics processing unit}
\newacronym{gpu}{GPU}{graphics processing unit}
\newacronym{hls}{HLS}{high-level synthesis}
\newacronym{hdl}{HDL}{hardware description language}
\newacronym{ims}{IMS}{iterative modulo scheduling}
\newacronym{isa}{ISA}{instruction set architecture}
\newacronym{ilp}{ILP}{integer linear programming}
\newacronym{ir}{IR}{intermediate representation} 
\newacronym{ii}{II}{iteration interval}
\newacronym{kms}{KMS}{kernel mobility schedule}
\newacronym{lut}{LUT}{lookup table}
\newacronym{mii}{mII}{minimum II}
\newacronym{ms}{MS}{mobility schedule}
\newacronym{msa}{MSA}{multiple sequence alignment}
\newacronym{pdu}{PDU}{power distribution unit}
\newacronym{pe}{PE}{processing element}
\newacronym{pl}{PL}{programmable logic}
\newacronym{ra}{RA}{register allocation}
\newacronym{rtl}{RTL}{register transfer level}
\newacronym{sat}{SAT}{satisfiability}
\newacronym{sm}{SM}{sequence matching}
\newacronym{soa}{SoA}{state of the art}
\newacronym{soc}{SoC}{system on chip}
\newacronym{tdp}{TDP}{thermal design power}
\newacronym{u}{U}{utilization}
\newacronym{wgs}{WGS}{whole genome sequencing}

\glsdisablehyper
\renewcommand*{\a}[1]{\gls{#1}}
\newcommand*{\as}[1]{\glspl{#1}}
\usepackage{textcomp}
\usepackage{balance}
\usepackage{placeins}

\usepackage[numbers,square]{natbib}
\usepackage[utf8]{inputenc}	
\usepackage[T1]{fontenc}	
\usepackage{xcolor}		
\usepackage[hyphens]{url}
\usepackage[colorlinks = false,
            hidelinks,
            linkcolor = purple,
            urlcolor  = blue,
            anchorcolor = black]{hyperref}	
\usepackage{booktabs} 		
\usepackage{nicefrac}		
\usepackage{microtype}		
\usepackage{lineno}		
\usepackage{float}			
\usepackage{tcolorbox}

\usepackage{newfloat}
\DeclareFloatingEnvironment[name={Supplementary Figure}]{suppfigure}
\usepackage{sidecap}
\sidecaptionvpos{figure}{c}

\usepackage{titlesec}
\titlespacing\section{0pt}{10pt plus 3pt minus 3pt}{1pt plus 1pt minus 1pt}
\titlespacing\subsection{0pt}{10pt plus 3pt minus 3pt}{1pt plus 1pt minus 1pt}
\titlespacing\subsubsection{0pt}{8pt plus 3pt minus 3pt}{1pt plus 1pt minus 1pt}
\titlespacing\section{0pt}{12pt plus 3pt minus 3pt}{1pt plus 1pt minus 1pt}
\titlespacing\subsection{0pt}{10pt plus 3pt minus 3pt}{1pt plus 1pt minus 1pt}
\titlespacing\subsubsection{0pt}{8pt plus 3pt minus 3pt}{1pt plus 1pt minus 1pt}

\usepackage{subcaption}
\usepackage{optidef}
\usepackage{multirow}
\usepackage[normalem]{ulem}

\usepackage{enumitem}
\setlist{nosep}  

\usepackage{tikz,hyperref}

\definecolor{lime}{HTML}{A6CE39}
\DeclareRobustCommand{\orcidicon}{
	\begin{tikzpicture}
	\draw[lime, fill=lime] (0,0)
	circle [radius=0.16]
	node[white] {{\fontfamily{qag}\selectfont \tiny ID}};
	\draw[white, fill=white] (-0.0625,0.095)
	circle [radius=0.007];
	\end{tikzpicture}
	\hspace{-3mm}
}
\foreach \x in {A, ..., Z}{\expandafter\xdef\csname orcid\x\endcsname{\noexpand\href{https://orcid.org/\csname orcidauthor\x\endcsname}
			{\noexpand\orcidicon}}
}

\newcommand{\TableMediumFont}{\footnotesize}

\newcommand{\hwname}{\textit{GeneTEK}}
\title{\hwname: Low-power, high-performance and scalable \\ FPGA architecture for exact unit-cost edit distance}
\date{}

\usepackage{authblk}

\author[1*]{Elena~Espinosa\orcidA{}}
\author[2+]{Rubén~Rodríguez~Álvarez\orcidB{}}
\author[3]{José~Miranda\orcidC{}}
\author[1,4]{\mbox{Rafael~Larrosa\orcidD{}}}
\author[5]{Miguel~Peón-Quirós\orcidE{}}
\author[1]{Óscar~Plata\orcidF{}}
\author[2]{\mbox{David~Atienza\orcidG{}}}

\affil[1]{Dept. of Computer Architecture, University of Malaga, Malaga, Spain}
\affil[2]{Embedded Systems Laboratory, École Polytechnique Fédérale de Lausanne (EPFL), Switzerland}
\affil[3]{Industrial Electronics Center (CEI), UPM, Madrid, Spain}
\affil[4]{Supercomputing and Bioinnovation Center (SCBI), University of Malaga, Malaga, Spain}
\affil[5]{EcoCloud, École Polytechnique Fédérale de Lausanne (EPFL), Switzerland}
\affil[*]{\tt elenamesga@uma.es}
\affil[+]{\tt ruben.rodriguezalvarez@epfl.ch}

\usepackage{titlesec}

\titleformat{\section}
  {\large\bfseries\raggedright\hyphenpenalty=10000}
  {\thesection}{1em}{}

\titleformat{\subsection}
  {\normalsize\bfseries\raggedright\hyphenpenalty=10000}
  {\thesubsection}{1em}{}

\titleformat{\subsubsection}[runin]
  {\normalsize\bfseries} 
  {\thesubsubsection.}
  {1em}
  {}
  [.]

\begin{document}

\newcommand{\keywords}[1]{%
  \noindent\textbf{Keywords: } #1
  \vspace{0.5em}
}

\twocolumn[
  \begin{@twocolumnfalse}

\maketitle

\vspace{-3em}
\begin{abstract}

The advent of next-generation sequencing (NGS) has revolutionized genomic research by enabling cost-effective, high-throughput sequencing of a diverse range of organisms. This breakthrough has unleashed a ``Cambrian explosion'' in genomic data volume and diversity. This volume of workloads places genomics among the top four big data challenges anticipated for this decade. In this context, pairwise sequence alignment represents a very time- and energy-intensive step in common bioinformatics pipelines. Speeding up this step requires the implementation of heuristic approaches, optimized algorithms, and/or hardware acceleration. Although state-of-the-art CPU and GPU implementations have demonstrated significant performance gains, recent FPGA implementations have shown improved energy efficiency. However, the latter often suffer from limited read-length scalability due to constraints on hardware resources when aligning longer sequences. In this work, we present a flexible FPGA-based accelerator template scalable up to 1000\,bp that implements Myers’s algorithm to compute exact unit-cost edit-distance using high-level synthesis and a worker-based architecture. GeneTEK, a set of instances of this accelerator template in a Xilinx Zynq UltraScale+ FPGA, achieves up to \SI{113}{\percent} increase in execution speed and up to 111$\times$ reduction in energy consumption compared to leading CPU and GPU solutions, while fitting comparison matrices up to 13$\times$ larger than previous FPGA-based systolic-array solutions. By following a SW--HW co-design approach, GeneTEK exploits parallelization at multiple levels and efficient memory use to deliver a scalable and accurate FPGA-based accelerator. These results reaffirm the potential of FPGAs as an energy-efficient platform for pairwise alignment of read-lengths up to 1000\,bp.

\end{abstract}
\vspace{0.35cm}

\keywords{
Energy efficiency, FPGA, genome assembly, high-level synthesis, Myers, reconfigurable architectures, scalable architectures, sequence alignment, sequence matching.
}

  \end{@twocolumnfalse}
]


\glsresetall
\section{Introduction}

Sequencing technologies enable the determination of an organism's genomic or transcriptomic sequences from small random fragments, known as reads. This capability is essential to study protein functions, the evolutionary history of species, personalized medicine, and more.


Since the completion of the Human Genome Project in 2003~\cite{international2001initial, international2001correction, international2004finishing}, the number of sequenced genomes has increased enormously~\cite{gibbs2020human}. Genomics data storage is estimated to rise to 40~exabytes per year, positioning it among the top four Big Data challenges of this decade~\cite{stephens2015big}.
Furthermore, data center concentration means that they account for \SI{20}{\percent} of electricity consumption in some countries, such as Ireland~\cite{iea2025}, while they are projected to account for up to \SI{3}{\percent} of global electricity consumption by 2026~\cite{iea2024}.
Consequently, there is a pressing economic, environmental, and social need to reduce the energy consumption associated with these computations.

This has led to a growing need for more efficient algorithms to process the massive amounts of data generated. In this context, pairwise sequence alignment continues to be the cornerstone of most bioinformatics pipelines and a primary source of computational load. One instance of this challenge is found in de novo genome assembly, specifically in overlap-layout-consensus (OLC) assembly~\cite{kececioglu1995exact, myers2005fragment, draghici2007systems, simpson2010efficient, simpson2012efficient, li2012comparison, rizzi2019overlap}, where the overlap step represents the major bottleneck since each read must be compared with every other read to identify overlaps~\cite{espinosa2023comparing}. This step is particularly challenging because most sequence pairs are not candidate matches, and the relative offset between sequences can be large and unknown. Genome assembly is fundamental since, although sequencing technologies provide the raw reads required for analysis, these fragments must first be accurately assembled to reconstruct the complete genome, a process that is performed de novo when no reference is available for comparison.

Other fields where pairwise alignment is essential include metagenomics—particularly read-based taxonomic profiling~\cite{quince2017shotgun}, where reads are assigned to reference genomes or marker genes—genetic variant detection studies, and \gls{msa}, which is required for domain analysis, phylogenetic reconstruction, and motif finding~\cite{chatzou2016multiple}.


Consequently, accelerating pairwise alignment is a high-priority goal in multiple areas of bioinformatics.

An efficient alternative to traditional pairwise sequence alignment algorithms is the \gls{bpm} algorithm, which uses bitwise operations to exploit data-level parallelism. In this way, the complexity is reduced from $\mathcal{O}(m \times n)$ to $\mathcal{O}(n)$, and the problem becomes suitable for hardware-based implementations, including \glspl{gpgpu} and domain-specific accelerators in \glspl{fpga}, both of which can deliver notable performance gains for large-scale bioinformatic pipelines~\cite{franke2020accelerating, jiang2021applications, o2023accelerating}. 

We find that reconfigurable hardware implementations are capable of outperforming high-end \gls{gpgpu} systems, especially from a power perspective~\cite{papadopoulos2013fpga, yano2014clast, robinson2021hardware}, even on low-cost \gls{fpga} platforms~\cite{fletcher2011bridging, fowers2012performance, papadopoulos2013fpga}. \mbox{\glspl{fpga}~\cite{xilinx1986}} are programmable logic with configurable logic blocks, providing fine- and coarse-grained parallelism. This capability enables faster execution and improved performance while reducing energy costs compared to traditional HPC systems~\cite{robinson2021hardware}.
In light of these advantages, we propose \mbox{\glspl{fpga}} as a flexible solution to solve the sequence alignment problem in data centers.

Specifically, in this work we propose a novel energy-efficient pairwise aligner using a single \gls{fpga} \gls{soc} platform that integrates the parallel Myers's bit-vector algorithm. The proposed architecture addresses the need for low-energy read mapping taking advantage of the massive parallelism of \glspl{fpga} at the sequence and bit-vector levels. The achieved implementation can be applied to OLC-based alignments with read lengths up to 1000\,bp.

We make four key contributions to the computational challenge of the genome alignment problem, namely:

\begin{enumerate}
    \item The design and implementation of an accelerated version of Myers's algorithm for pairwise sequence alignment on an \a{fpga}
    that targets up to 1000\,bp read lengths, surpassing the scalability of feasible read lengths sizes of previous high-throughput FPGA designs.\footnote{Throughout the paper, we define the scalability of an FPGA design as the feasibility of targeting larger problem sizes, which is defined by the read length.}~Moreover, the design is able to maintain higher performance than other CPU and GPU state-of-the-art solutions within this range.
    Our design leverages two levels of parallelism: wide-bit data representation for explicit vectorization of sequences, and a flexible parallelization of multiple pair comparisons that enables the design to maximize resource usage of a target \a{fpga}. Specifically, we achieve a maintained high-throughput scaling for reads in the \num{200}--\num{500}\,bp range, which includes typical lengths for widely used sequencing technologies such as Illumina~\cite{espinosa2024advancements}, including recent 500\,bp options.\footnote{\url{https://www.illumina.com/systems/sequencing-platforms.html}}
    
    \item Evidence of the compute-boundness of Myers's algorithm when correctly dimensioning the design's computing and buffering elements. Our \gls{fpga} design exploits internal memory resources to cache sequences and drastically reduce the volume of data read from the system RAM.
    
    \item The proposal of an \a{fpga}-based SW-HW co-design approach using \a{hls} to develop a template-based accelerator that is parameterizable at design time. The template is used to instantiate an accelerator optimized for the size of the problem and the resources of the platform. Our proposal optimizes the resource usage of \as{fpga} by adjusting the number of parallel pair alignment units, the buffer size for sequences, and the maximum sequence length. This co-design approach is easy to integrate into other bioinformatics pipelines by deploying the complete application on a Linux operating system in the \a{fpga}.
    
    \item A comprehensive evaluation of the energy efficiency of state-of-the-art pairwise alignment implementations.
    Given the scarcity of previous analyses on energy efficiency, we measure the energy efficiency of state-of-the-art CPU, GPU, and \gls{fpga} implementations.
\end{enumerate}
 
The remainder of this paper is structured as follows. Section~II provides a detailed review of Myers's algorithm. Section~III discusses related work in the field. Section~IV outlines our implementation methods. Section~V presents the evaluation methodology. Section~VI presents the results of our findings. Finally, in Section~VII, we draw our conclusions.


\section{The read mapping problem and Myers's algorithm}

Sequence alignment can be defined as a problem of \gls{sm}, where the aim is to compare one text with another, both of which represent the genetic alphabet (adenine (A), guanine (G), cytosine (C) and thymine (T) for DNA, and uracil (U) for RNA) to find equalities, dissimilarities or occurrences of this pattern in the text. 

In bioinformatics, \gls{sm} problems can be solved using classical \gls{dp} algorithms, such as the Levenshtein distance~\cite{levenshtein1966binary}, Needleman-Wunsch~\cite{needleman1970general}, Smith-Waterman~\cite{smith1981identification} and Smith-Waterman-Gotoh (SWG)~\cite{gotoh1982improved}. However, these proposals result in quadratic time and space complexity, that is, $\mathcal{O}(m \times n)$ between two sequences of lengths $m$ and $n$. One candidate to replace the classical \gls{dp} algorithms is the \gls{bpm} algorithm~\cite{myers1999fast}.\looseness=-1000

Myers introduces improvements over the traditional re\-source-in\-ten\-sive \gls{dp} matrix under Levenshtein edit distance~\cite{levenshtein1966binary} by re-encoding the matrix and processing it column-by-column using bit-vectors. This proposal achieves a linear runtime of $\mathcal{O}(n)$ when using no more than 32 or 64 bits, where $n$ represents the length of the text. This inherent data parallelism enables effective use of the parallel processing capabilities of \glspl{gpgpu} and \glspl{fpga}.

In particular, \gls{sm} aims to identify differences and similarities between two sequences. Let $Q$ (query) and $P$ (target) be two strings of elements from the genetic alphabet
with lengths $|Q|=m$ and $|P|=n$; \gls{asm} tries to find all the locations in $P$ where the distance $E$ to $Q$ is at most $k$ differences. These differences, denoted by edits, can be classified as substitutions, deletions, or insertions in one or both sequences. The cumulative cost of these edits represents the edit distance.

\begin{figure}[ht!]
    \centering
 \includegraphics[width=\columnwidth]{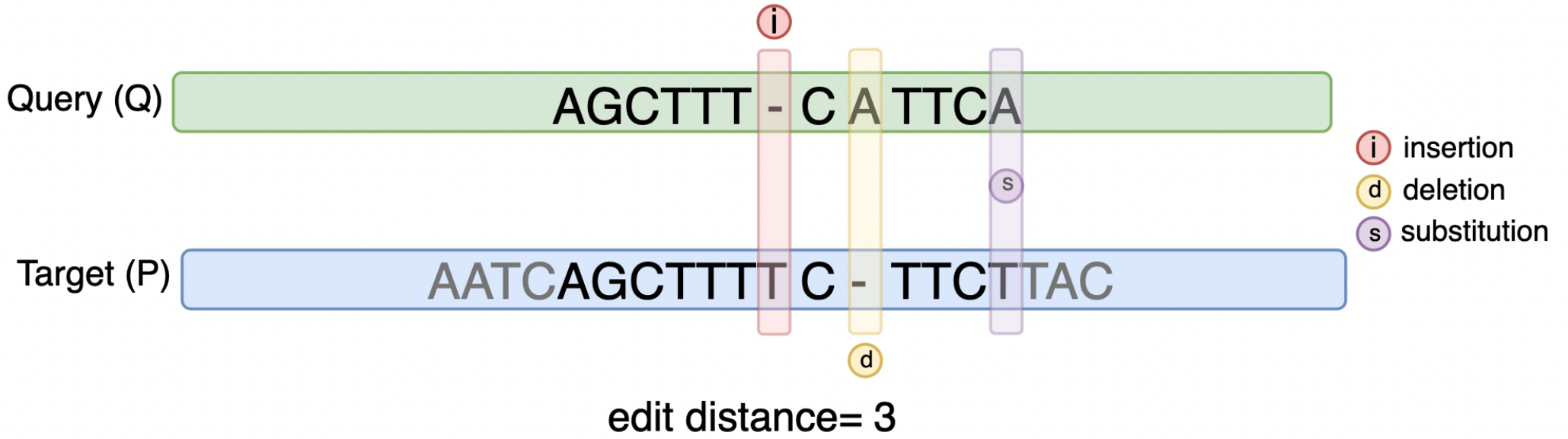}
    \caption{Three types of errors (i.e., edits).}
    \label{fig:Fig1}
\end{figure}

The most relevant so-called edit distances in biotechnology are: Hamming distance~\cite{hamming1950error}, Levenshtein~\cite{levenshtein1966binary} distance and Damerau-Levenshtein~\cite{damerau1964technique} distance. Hamming distance considers only substitutions and applies only to strings of equal length. In contrast, the Levenshtein distance accounts for insertions and deletions as well, and the Damerau-Levenshtein distance further includes transpositions. Figure~\ref{fig:Fig1} shows each possible kind of edit. Whereas determining the Hamming distance is a relatively easy task and has been effectively addressed, the computation of the Levenshtein and Damerau-Levenshtein distances involves a higher computational cost and remains a challenge when dealing with data-intensive applications.  

Myers's algorithm~\cite{myers1999fast} addresses the approximate string-matching problem by computing the Levenshtein distance between two texts, allowing for a maximum of $k$ errors using minimal memory and a few bit operations.

The main idea is to parallelize the DP matrix computing the column as a whole in a series of bit-level operations. Myers re-encoded the DP scoring matrix by accounting only for the vertical and horizontal delta values among the adjacent cells. The result is four bit-vectors that represent four states in each column $j$ in the matrix, with $j$ in $1...n$: horizontal positive ($HP_j$), horizontal negative ($HN_j$), vertical positive ($VP_j$), and vertical negative ($VN_j$). These bit-vectors are used to compute each column by performing specific bitwise logical, shift, and addition operations based on the bit-vectors of the preceding column. Based on these ideas, the scoring matrix is obtained by sequentially calculating the bit-vector states and determining the scores at the bottom row. 
To compute the edit distance, the algorithm starts with the maximum distance in the lower left cell and increases or decreases it by $\pm$1 in each iteration based on the last bits of $HP_j$ and $HN_j$. In particular, the distance increases by 1 if $HP_j$ is set and decreases by 1 if $HN_j$ is set. Both bits cannot be set simultaneously. However, it is possible that neither bit is set, which indicates that there is no change in distance between them.

Algorithm~\ref{alg:algorithm1} shows the pseudo-code for Myers’s algorithm, which follows these steps: 
\begin{enumerate}
    \item Preprocess the variables for column 0 (lines 5--9).
    \begin{itemize}
        \item  Set the $Peq$ array for the $Query$ sequence (line 6). The bit-vectors or bitmask values are calculated for each nucleotide in the target sequence (referred to as the $Peq$ vector). %
        This vector represents the presence of each nucleotide in the target sequence at each position in the query sequence, e.g., given the query ``AATC'', the bit-mask for the nucleotide `A' is 1100.
        \item Initialize the bit-vectors $VP$ and $VN$ (lines 8--9).
    \end{itemize}
    \item Scan and compute the DP matrix from left to right by column (lines 11--27).
    \begin{itemize}
        \item Compute $HP$, $HN$ and $D0$ of column $j$ from $VP$ and $VN$ of column $j-1$ (lines 13--15).
        \item Compute $VP$ and $VN$ of next column using $HP$, $HN$ and $D0$ (lines 17--18).
        \item Estimate the score from $HP$ and $HN$ (lines 19--23).
    \end{itemize}
    \item Output the locations with the lowest edit distance as the optimal end location on the target sequence (line 28).
\end{enumerate}

The complexity of Myers's algorithm is $\mathcal{O}(n \lceil m/w \rceil)$, where $w$ represents the word size of the machine (e.g., 32 or 64 bits). By packaging queries in units of up to 64 bits, such as using \texttt{unsigned long long int}, the overall time requirements are reduced to $\mathcal{O}(n)$.
However, reads of 65 nucleotides or \glspl{bp} and higher require multiple CPU registers to emulate bit-vectors and perform operations, which leads to degraded performance due to additional processing overhead. 

\begin{figure}[ht!]
    \centering
 \includegraphics[width=\columnwidth]{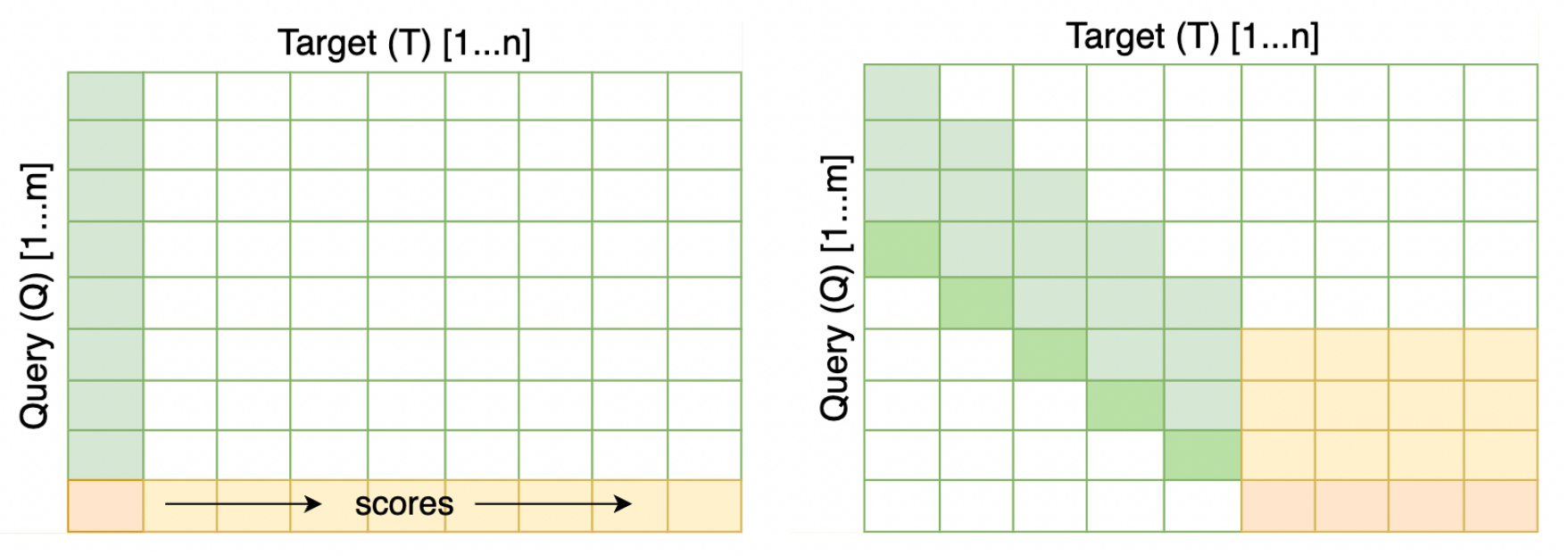}
    \caption{Myers's computing matrix (left) and banded algorithm scheme (right) for a band width of 4 nucleotides.}
    \label{fig:Fig2}
\end{figure}

Hyyrö~\cite{hyyro2002faster} addressed this issue by reducing the number of nucleotides processed in each column and adopted a \textit{banded} strategy to compute the matrix, since it is not necessary to consider more than \SI{10}{\percent} to \SI{20}{\percent} of the query length as an acceptable difference rate for finding similarities. The main concept relied on the use of bit-vectors that correspond only to the width of the \textit{band}, rather than the requirement of bit-vectors that match the length of the query. Figure~\ref{fig:Fig2} shows Myers's computing matrix (left) and the band-based version of Hyyrö's proposal (right).
However, the main disadvantage of this proposal is that the search space in the target sequence is constrained. Thus, if the alignment regions are located at distant positions between the two sequences, accurate identification is hindered. 
For instance, the banded approach becomes a limitation in all-vs-all comparisons where the relative offset between two sequences can be large and unknown, potentially requiring wide bands or prefiltering. Conversely, when candidate pairs are well-localized, banding can be highly efficient.

This makes Myers's algorithm a more precise solution than Hyyrö's proposal for the sequence alignment problem, although additional strategies are required to fully exploit its inherent data-level parallelism.

\begin{algorithm}[!ht]
\caption{Myers’s bit-vector proposal~\cite{myers1999fast}}\label{alg:algorithm1}
\begin{algorithmic}[1] 
\footnotesize
\Function{Myers}{Q, P, m, n, k} 
\Comment{$P$ and $Q$: target and query respectively; $n$ and $m$: size of the target and query respectively; $k$: defined threshold}
\For { $\sigma$ $\in$ all characters} \Comment{For genome sequencing, all characters = \{A,T,C,G\}}
\State $Peq_\sigma \leftarrow 0 $;
\EndFor
\For{$i$  $from$ $1$ $to$ $m$}\Comment{Preprocessing}
\State $Peq_{Q[i]} \leftarrow Peq_{Q[i]} \verb\|\  0^{m-1}10^{i-1}$;
\EndFor 
\State $VP\leftarrow1^m$;
\State $VN\leftarrow0^m$;
\State $score \leftarrow m$;
\For{$j$  $from$ $1$ $to$ $n$}\Comment{Searching}
    \State $X \leftarrow Peq_{P[j]}$ \verb\|\ $VN$;
    \State $D0 \leftarrow ((VP + (X$ \verb\&\ $VP))$ \verb|^| $VP)$ \verb\|\ $X$; \Comment{Diagonal zero delta vector}
    \State $HN \leftarrow VP$ \verb\&\ $D0$; \Comment{Horizontal negative delta vector}
    \State $HP \leftarrow VN $ \verb\|\ $\verb|~|(VP$ \verb\|\ $D0)$; \Comment{horizontal positive delta vector}
    \State $X \leftarrow HP$  \verb|<<| $1$; \Comment{Shifting HP for the next iteration}
    \State $VN \leftarrow X$ \verb\&\ $D0$; \Comment{Vertical negative delta vector}
    \State $VP \leftarrow (HN$ \verb|<<| $1)$ \verb\|\ $\verb|~|(X$ \verb\|\ $D0)$; \Comment{Vertical positive delta vector}
    \If{$HP $ {\tt \&} $10^{m-1}$}  \Comment{Scoring} 
        \State $score \leftarrow score + 1$;
    \ElsIf{$HN$ {\tt \&} $10^{m-1}$}
        \State $score \leftarrow score - 1$;
    \EndIf
    \If{$score \leq k$}  
        \State Report a match ending at $P_j$;
    \EndIf
\EndFor
\State \Return $score$;
\EndFunction
\end{algorithmic}
\end{algorithm}


\section{Related work}

Several designs have been proposed in the literature to accelerate the alignment process and enhance computational efficiency in specific environments. In particular, Myers's bit-vector algorithm is a widely used algorithm for efficient \gls{asm}. This \gls{dp}-based algorithm calculates the edit distance between two strings using addition, shifting, and bitwise operations. Therefore, numerous enhancements and optimizations based on SIMD instructions, \glspl{gpu}, GPUs+CPUs, and \glspl{fpga} have been proposed in the literature for both Myers's algorithm and the banded version of Hyyrö, particularly in the context of short-read sequence alignment.

Targeting CPUs, \textit{Edlib}~\cite{vsovsic2017edlib} is the first known implementation of Myers's algorithm and is used by \textit{Medaka}~\cite{medaka} and \textit{Dysgu}~\cite{cleal2022dysgu}. It uses Ukkonen's banded algorithm to reduce the space of search and extends Myers's algorithm to support the global and the semiglobal alignments. Moreover, the \textit{SeqAn} library~\cite{doring2008seqan} implements Myers's algorithm using SIMD instructions. Furthermore, the state-of-the-art provides multiple CPU implementations for sequence alignment, including SIMD libraries such as \textit{Parasail}~\cite{daily2016parasail} and \textit{KSW2}~\cite{suzuki2018introducing} (implemented in \textit{minimap2}~\cite{li2018minimap2}), as well as other efficient libraries, such as \textit{WFA2-lib}~\cite{marco2021fast}. \textit{WFA2-lib} implements the wavefront alignment (WFA) algorithm using simple computational patterns that can automatically vectorize across different architectures without requiring code adaptation.

\Glspl{gpu} have also been widely adopted as hardware accelerators in multiple bioinformatics applications because they provide higher computational throughput and memory bandwidth compared to traditional multi-core processors. Thus, several \gls{gpu} implementations are available, such as Chacón~\cite{chacon2014thread}, which optimizes a CUDA version of Myers's algorithm using a thread-cooperative approach. Additional \gls{gpu}-based libraries like \textit{GASAL2}~\cite{ahmed2019gasal2} provide efficient implementations of different sequence alignment algorithms, including \textit{Needleman-Wunsch} and \textit{Smith-Waterman}. Others, such as \textit{WFA-GPU}\cite{aguado2023wfa}, present a CPU-GPU co-design capable of performing inter-sequence and intra-sequence parallel sequence alignment. Similarly, libraries such as \textit{Scrooge} implement a fast genomic sequence aligner designed for CPUs, \glspl{gpu}, and \glspl{asic}. Specifically, \textit{Scrooge} outlines the limitations of \textit{GenASM}~\cite{cali2020genasm} and proposes a fast, memory-efficient genomic sequence aligner that outperforms both the CPU and \gls{gpu} implementations of \textit{GenASM}.

Many of these existing solutions implement gap-affine scoring, which incurs higher computational cost and is therefore well suited for alignment refinement, but less suitable for the initial search over large alignment volumes. In addition, some GPU implementations rely on WFA-based methods, whose computational cost increases with edit distance. These algorithms typically exhibit reduced throughput in settings where there is no prior knowledge of matching candidates (e.g., exhaustive all-vs-all comparisons).

The high cost and energy consumption of \gls{gpu}-based clusters often make \glspl{fpga} more suitable for genomic data processing, since they offer lower cost and higher energy efficiency while maintaining massive parallelism capabilities.
Within this context, L. Cai~\cite{cai2019design}, D. P. Bautista~\cite{bautista2021bit}, and Rufas~\cite{castells2021opencl} propose \gls{fpga}-based implementations of Myers's algorithm. Among them, Rufas's implementation outperforms the others, although it employs a band heuristic based on Hyyrö's algorithm, which compromises the accuracy of the alignment. However, none of these implementations addresses energy consumption.

A more recent approach by \textit{Schifano}~\cite{schifano2025high} surpasses even the previous solution of Rufas in terms of performance and addresses energy consumption. Despite these improvements, \textit{Schifano}'s implementation still presents drawbacks. First, like Rufas's implementation, it relies on a banded approach to process reads longer than \num{224}~nucleotides. This banded approach limits the matching accuracy when no prefiltering or prior localization is available. Second, the authors do not demonstrate the effectiveness of their solution with variable-length reads, leaving its performance under different read sizes uncertain.
This is relevant because among widely used sequencing technologies, only Illumina generates fixed-length reads, and even these reads can become variable in length after trimming (e.g., removing adapters or low-quality bases). Fixed-length processing also excludes studies that require mapping of sequencing reads against marker genes or for homology analysis.\looseness=-1

Other recent proposals, such as Venkat Gudur~\cite{gudur2021fpga}, implement an energy-efficient algorithm based on Myers's to accelerate the mapping of many sequencing reads to a single very long sequence, which is the reference genome. However, while this approach resolves reference-guided alignment, it does not address large-scale many-to-many pairwise alignment. Thus, accelerating such workloads requires dedicated high-performance strategies capable of handling large data volumes.

To address these limitations, we introduce \hwname{}, a fast and energy-efficient \gls{fpga} accelerator instance based on a SW-HW co-design approach using \gls{hls}. In contrast to previous work, we 1) implement an exact pairwise sequence aligner that does not rely on banded heuristics and 2) provide a comprehensive evaluation of both performance, energy, and memory consumption.


\section{\hwname{}}

This section provides an architectural overview, implementation methodology, and execution flow of \hwname{}.
\hwname{} uses the internal memory resources of the \a{fpga} to reduce data traffic with the external system memory, and the extensive logic resources of the \gls{fpga} to exploit parallelism at two levels: at the task level, using independent workers, and at the bit-vector level, using Myers's algorithm inside each worker.

\subsection{Architecture overview}\label{subsec:architecture}

\Glspl{fpga} include several types of configurable resources on their \gls{pl} side, such as \glspl{lut}, \glspl{ff}, \glspl{dsp}, and \glspl{bram}.
An \gls{fpga} \gls{soc} may also include additional hard-IP blocks that connect to the \a{pl}, such as ARM cores. Additionally, the \gls{soc} is usually paired with \gls{dram}, which serves as the system's main memory.
\as{fpga} achieve high efficiency by tailoring the configuration of the internal resources and their interconnection---according to a HW design---for a particular task.
In this way, lower power \as{fpga} can match the performance of more power-hungry CPUs and \glspl{gpu}, ultimately reducing the overall energy consumption during computation.

\begin{figure}[ht]
    \centering
 \includegraphics[width=\columnwidth,trim={0.5cm 0 0.5cm 2.5cm},clip]{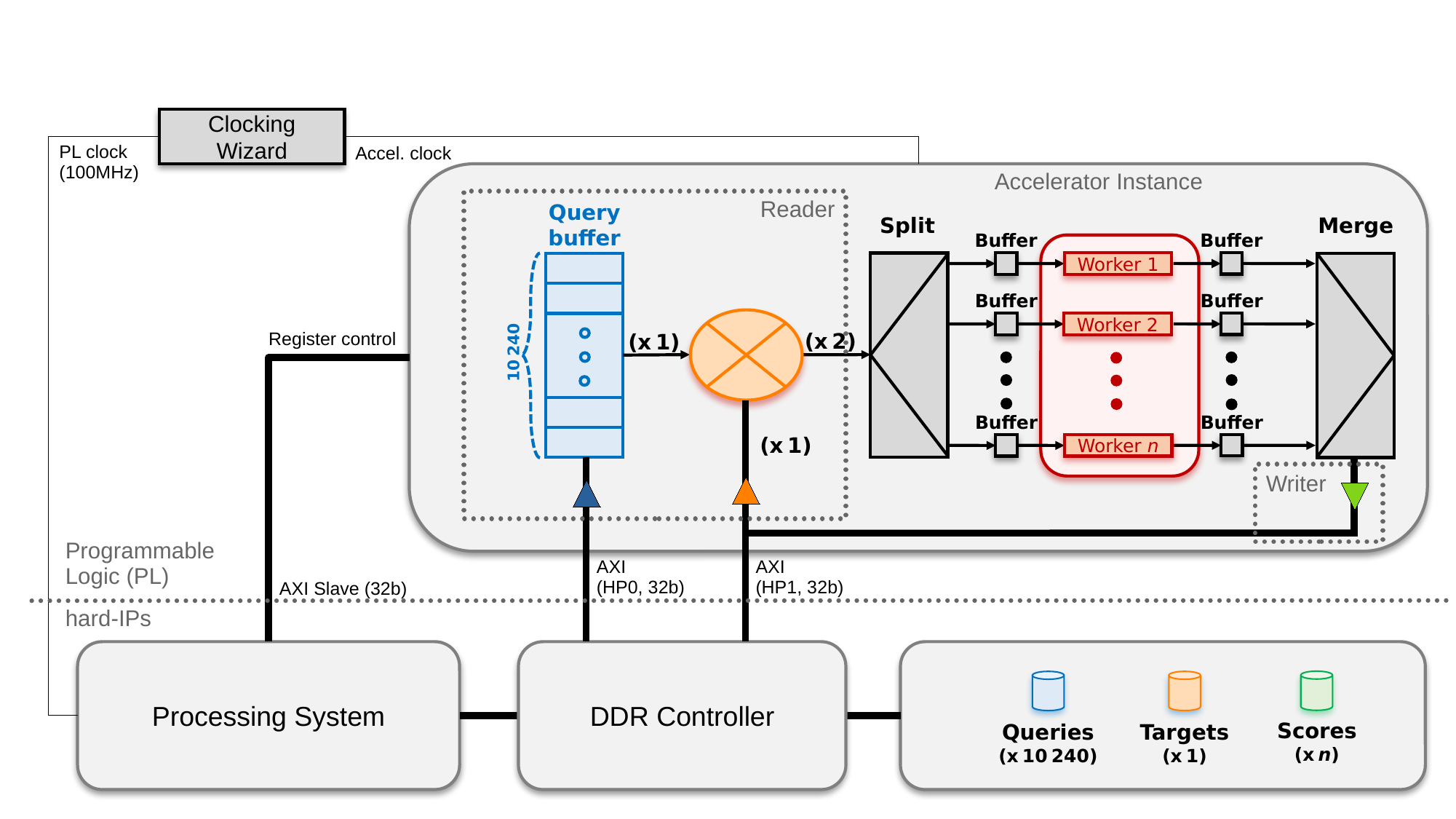}
    \caption{Architecture of \hwname{}. To reduce memory accesses, \hwname{} implements a query buffer. Target sequences are read one by one and compared against the sequences in the query buffer, sending query-target pairs to the workers. Once all the targets have been compared with the queries in the buffer, \hwname{} reads a new set of queries into the internal buffer and repeats the process for all the targets.}
    \label{fig:gsystemArchitecture}
\end{figure}

Figure~\ref{fig:gsystemArchitecture} illustrates the architecture of \hwname{}, implemented as a hardware accelerator in the \gls{pl} and coupled to a processing system that manages memory allocation and execution.
The accelerator targets the \gls{sm} problem, which performs pairwise alignment between a set of query sequences and a stream of target sequences.

To avoid memory-bound execution, \hwname{} adopts a buffered streaming architecture in which queries are cached in on-chip \glspl{bram} and reused across multiple target comparisons, while target sequences are streamed from off-chip memory.
A Reader stage loads queries and targets, a Split stage distributes query--target pairs to $W$ parallel Workers in a round-robin fashion, and each Worker implements a pipelined Myers's alignment kernel (with a query-size level of parallelism).
The results are collected by a Merge stage and streamed back to memory by the Writer via AXI interfaces.
The accelerator presents a set of control registers through an AXI lite slave interface.

This organization minimizes off-chip memory traffic, enables scalable parallelism through parameterized worker replication, and sustains high throughput across platforms, with all pipeline stages operating at an initiation interval of $\mathrm{II}=1$ so that no stage limits steady-state performance.

\begin{table}[tp]
    \caption{Reductions in data movement volume and time obtained by \hwname{}, for \num{e6} query sequences and \num{e6} target sequences with an average string length of \num{100}~nucleotides. Transfer times with a 64-bit DDR4~3200~DRAM}
    \centering
    \TableMediumFont{
    \begin{tabular}{lr@{\hskip0.25cm}r@{\hskip0.25cm}r@{\hskip0.25cm}}
    \toprule
     & Buffer size & Data transferred & Transfer time\\
     & (sequences) & (\si{\gibi\byte}) & (\si{\second})\\ \midrule
    No buffering & -- & \num{93132.4} & \num{3906.3}\\ \midrule
    \hwname{} & \num{1024} & \num{91.0} & \num{3.8}\\
    \hwname{} & \num{10240} & \num{9.2} & \num{0.4}\\
    \bottomrule
    \end{tabular}}
    \label{tab:bufferingImpact}
\end{table}

The first optimization introduced in \hwname{} is to use internal \gls{fpga} memory to cache strings.
Unfortunately, memory resources are limited; for example, our target \gls{fpga} has \SI{4.75}{\mebi\byte} of internal SRAM storage.
On the upside, these memories are very energy-efficient and they typically require just one clock cycle to read a complete string.
In \hwname{}, we create a buffer of \num{10240}~entries for query sequences and, hence, the \gls{sm} process is divided in chunks of \num{10240}~queries.
Therefore, \hwname{} first reads \num{10240}~queries and stores them locally in the \gls{fpga} buffer; then, it reads the target sequences one by one, performing each time the comparison of that one target sequence against all the query sequences stored in the buffer.
As shown in Table~\ref{tab:bufferingImpact}, this buffering scheme speeds up the complete process 
and removes memory bandwidth as a bottleneck.

As a result, buffering \num{10240}~queries shifts the accelerator’s roofline model~\cite{j:rooflinePatterson} to the right, enabling an increase of up to four orders of magnitude in the input computational intensity, defined as
\begin{equation}\label{eq:oi_in}
    \mathit{OI}_{\mathit{in}} =
    \frac{N_{q} \times \overline{L_{q}} \times N_{t} \times \overline{L_{t}}}
    {N_{q} \times \overline{L_{q}} + \left\lceil N_{q}/B_{q} \right\rceil \times (N_{t} \times \overline{L_{t}})} ,
\end{equation}
where $\overline{L_{q}}$ denotes the average query length in base pairs, $N_{q}$ is the number of queries in the dataset, $N_{t}$ is the number of target sequences, $\overline{L_{t}}$ is the average target length, and $B_q$ is the query buffer size.
In contrast, the output operational intensity defined by
\begin{equation}\label{eq:oi_out}
    \mathit{OI}_{\mathit{out}} = \frac{\overline{L_{q}} \times \overline{L_{t}}}{4}.
\end{equation}
remains fixed and is determined by the 32\,bit word used to write the alignment scores.
Finally, to increase the information density in the internal \gls{fpga} memories and reduce the bitwidth of the datapaths, the accelerator encodes each nucleotide internally using a two-bit binary representation: 00 for adenine (A), 01 for cytosine (C), 10 for thymine (T), and 11 for guanine (G).

The second optimization introduced in \hwname{} is the parallelization of the comparison of query-target pairs.
The design includes a configurable number of workers, which can be increased based on the resources available in the specific \gls{fpga}.
The \gls{sm} tasks are parallelized in all workers.
In its current implementation, \hwname{} can send one pair of sequences to one worker per cycle (owing to the capacity to read one complete query sequence from the internal buffer in one cycle).
Upon completion, each worker sends its result to a writer module along with the index of the comparison, so that the results can be written back to system memory in the right positions.
By employing several workers, each of them handling one individual \gls{sm} operation at a time, the workload can be distributed over different \gls{fpga} resources.

Finally, \hwname{} exploits the bit-level parallelism of Myers's algorithm, which is pivotal for \gls{sm} operations.
This algorithm leverages bit manipulation to enhance the speed of matching operations, enabling linear scalability of the processing time with the length of the string.
Each worker utilizes this algorithm internally to compare a pair of strings; the workers build one column of the \gls{dp} matrix per cycle---processing in parallel all the characters.

\begin{table*}[tp]
    \caption{Synthesis time and energy consumption. \hwname{} short synthesis time enables agile design cycles. For~\cite{schifano2025high}, we use the numbers reported for solution \num{10} in Table~2, which is the closest to ours in terms of reads length. Average power during synthesis of \SI{230}{\watt} (measured with the PDUs of the EcoCloud experimental data center\protect\footnotemark)}
    \centering
    \begin{tabular}{lr@{\hskip0.25cm}r@{\hskip0.25cm}rr}
    \toprule
    Work & HLS time & Vivado time & Total time & Energy-to-design\\
     & (\si{\second}) & (\si{\second}) & (\si{\second}) &  (\si{\joule}) \\ \midrule
    \hwname{} & \num{89} & \num{4929} & \num{5018} & \num{1154140}\\
    \cite{schifano2025high} & \num{1744134} & \num{261492} & \num{2005626} & \num{461293980}\\
    \bottomrule
    \end{tabular}
    \label{tab:designTimeAndEnergy}
\end{table*}

By combining these two levels of parallelism, the design maximizes resource utilization and increases performance.
The parallel workers ensure coarse-grain data and task parallelism, while the internal bit-vector operations ensure that each worker executes string matching as efficiently as possible.
This architecture leads to FPGA solutions for \gls{sm} applications with larger read-length scalability.
For example, compared to approaches based on systolic arrays~\cite{schifano2025high}, whose area scales quadratically with the length of the strings, \hwname{}---based on independent workers that exploit bit-level parallelism---scales linearly in area with the string length and the number of workers, thus allowing future designs to process larger reads produced with modern technologies and expand to larger \glspl{fpga}.

\subsection{Template-based implementation methodology}

In this section, we explain how \hwname{} is designed using \gls{hls} by describing its architecture as a parametrizable template.
The higher level of abstraction offered by the \gls{hls} tools makes the description of the HW easier to maintain or modify and results in increased productivity.
However, it may produce less efficient HW compared to designs implemented using \glspl{hdl}.
In this work, we show that an \gls{hls} approach is capable of achieving improvements in both latency and energy for the \gls{sm} problem, without sacrificing productivity.

The \hwname{} template is defined by a combination of design-time parameters, including the maximum supported sequence length $L_{\max}$, the query buffer size $B_q$, the number of parallel workers $W$, the target clock frequency for Vitis HLS synthesis (which guides scheduling and pipelining decisions), and the target clock frequency of the Vivado clocking wizard used for final deployment (A \SI{100}{\mega\hertz} base clock is provided to the PL clocking wizard). The selection of these variables serves two purposes: adapting the design to different \glspl{fpga}, thereby increasing the level of parallelization based on available resources, and configuring the maximum string length to be processed to the increasing read lengths produced by new technologies.

The accelerator template uses \gls{hls} task structures to implement independent tasks, which are interconnected through \gls{fifo} streams.
The tasks start when there are data available in the input \gls{fifo} and space to write in the output \gls{fifo}. This task-based pipelined design enables better concurrency natively.

The template uses \gls{hls} wide-bit data formats to efficiently represent strings,
which allows the \a{hls} interpreter to handle automatically advanced bit manipulation and arithmetic operations.
This streamlines the development process and optimizes performance, as the \a{hls} tool can generate efficient hardware tailored to the specific operations and bitwidths needed.
The tool automatically stores these wide-bit representations interleaved across multiple \glspl{bram} and \glspl{ff} in the design.
This interleaving enables the parallel access to data required in Myers's algorithm, significantly improving throughput and reducing bottlenecks associated with memory accesses, since a complete string can be accessed in a single cycle.

As a result, the internal \gls{fifo} streams to the workers are configured with a width of $4 \times L_{\max} + 2 \times \lceil \log_2(L_{\max}) \rceil + 32$ bits to encode the two sequences, their respective lengths, and the alignment position identifier. Similarly, the output \glspl{fifo} of the worker is configured with a width of $\lceil 2 \times \log_2(L_{\max}) \rceil + 32$ bits to encode the score and the position identifier. All intermediate \glspl{fifo} for the streams are configured with the default depth of two elements.

On the memory side, queries are stored using a wide-bit representation that enables parallel access to the entire sequence in a single cycle.
For example, a 360\,bp query encoded with 2\,bits per base requires 720\,bits, which is automatically mapped by the \gls{hls} compiler onto 12~parallel \gls{bram} blocks each configured for 64-bit access ($360\,\mathrm{bp} \times 2\,\mathrm{bits} / 64\,\mathrm{bits} = 11.25$ \gls{bram} words) to sustain single-cycle access.

Each pair of strings is distributed independently to a worker unit, which maximizes computational efficiency and avoids memory collisions, allowing each worker to perform its task asynchronously without interference from other workers.
Although this design choice replicates the reference string---one for each worker---it does not affect the feasible design size, since it is limited by logic resources (\glspl{lut}) rather than by memory resources (\glspl{bram} and \glspl{ff}).
The next likely bottleneck in this design is the scheduler that sends tasks to workers.
Possible solutions include the division of the queries buffer to accommodate multiple independent schedulers, each of them sending tasks in parallel to disjoint sets of workers.

Overall, our approach creates a highly efficient and scalable framework for processing string data that accommodates varying data sizes, string lengths, and the specifications of the targeted \gls{fpga} hardware.
As an additional benefit, Table~\ref{tab:designTimeAndEnergy} shows that this worker-based approach generates designs that are easier to process for the \gls{hls} tools, which reduces the energy consumed during the design phase and unlocks much faster design cycles.\footnotetext{https://ecocloud.epfl.ch/2024/05/15/our-new-experimental-facility-is-up-and-running/}\looseness=-1

\subsection{Execution flow}

The end-to-end execution flow of \hwname{} consists of a program that runs on the cores of the \a{fpga} SoC.
As a first step, the target and query sequences are loaded into the system memory in ASCII format. The program then configures \hwname{}'s registers for the \gls{sm} tasks ahead.\looseness=-1

The accelerator translates each nucleotide in the sequences from ASCII encoding into the aforementioned two-bit representation before storing them in the internal memories. After the computation is completed, the results are written back to system memory. A key benefit of this architecture is that while the accelerator is busy with these tasks, the cores can be freed to run other operations or, in scenarios where power conservation is a priority, enter a low-energy sleep mode, hence enhancing overall system efficiency and resource management.

For datasets that exceed the system \gls{dram} size, \hwname{} implements ping-pong buffering. This method enables efficient parallelization of storage-to-memory transfers and processing with the accelerator.


\section{Evaluation methodology}

\subsection{Input data}\label{dataset_description}

We perform pairwise alignments between all reads in simulated datasets. To cover different use cases in bioinformatics, we generate datasets with reads of uniform (fixed) length and reads of variable length.
We use fixed-length reads to emulate alignments from uniform technologies like Illumina in scenarios such as OLC-based assembly with read lengths up to 1000\,bp, and variable-length reads to reflect two key use cases: 1) mapping reads against specific sequences, such as marker genes, and 2) performing alignments in homology studies, variant analysis, and other genomic investigations.

To generate the datasets, we use the next-generation simulator \textit{NGSNGS}~\cite{ngsngs2023} to generate a total of 28 datasets, organized into two groups: \textit{Group A} consists of 16 datasets, each combining one of seven fixed read lengths (\num{100}\,\glspl{bp}, \num{200}\,\glspl{bp}, \num{260}\,\glspl{bp}, \num{300}\,\glspl{bp}, \num{360}\,\glspl{bp}, \num{500}\,\glspl{bp} or \num{1000}\,\glspl{bp}) with one of four sample sizes (\num{1000}, \num{5000}, \num{10000}, or \num{100000} reads), and \textit{Group B}, which includes 12 datasets resulting from combinations of variable read lengths within three ranges (\num{100}--\num{160}\,\glspl{bp}, \num{200}--\num{260}\,\glspl{bp}, and \num{300}--\num{360}\,\glspl{bp}) and four sample sizes (\num{1000}, \num{5000}, \num{10000}, and \num{100000} reads).

\begin{figure}[ht]
    \centering
 \includegraphics[width=\columnwidth]{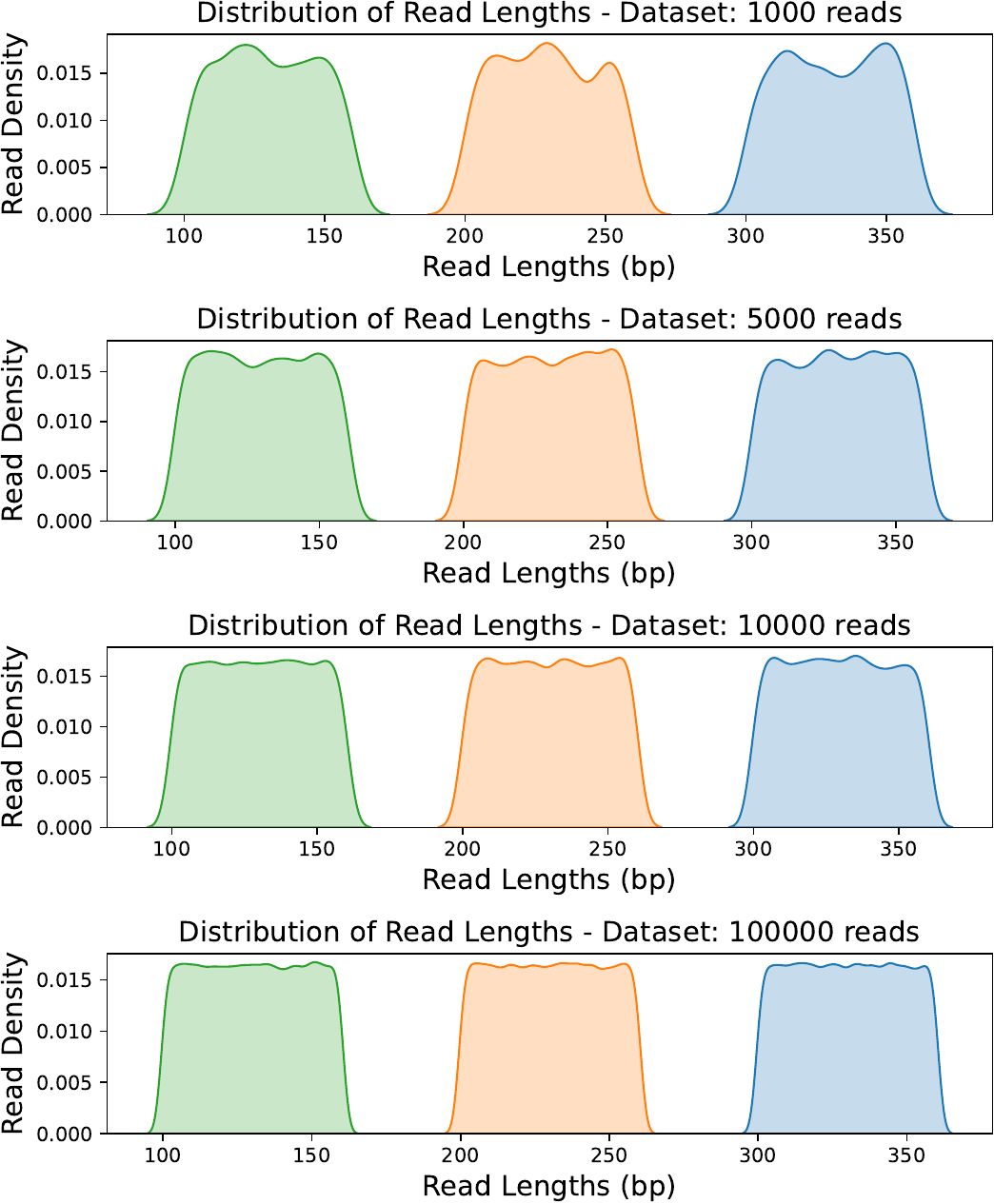}
    \caption{Distribution of read lengths for each simulated FASTQ file.}
    \label{Figure4}
\end{figure}

To generate each set of reads, we select random sequences from the reference genome of \textit{Pan troglodytes}~\cite{PanTroglodytesGenome}. We estimate the length of the random sequences based on the dataset size and the read length (for datasets with fixed-length reads) or the average read length (for datasets with variable-length reads) to ensure a coverage depth of at least 40$\times$. Theoretically, high coverage is critical for \emph{de novo} assembly in the pairwise alignment step because it guarantees that each position in the genome is sequenced multiple times, which increases the likelihood that any two reads will share a true overlap. In practice, ensuring high coverage in our evaluations guarantees a sufficient number of overlapping reads to enable reliable pairwise alignments.
Figure~\ref{Figure4} illustrates the distribution of read lengths in the generated FASTQ files for the simulated datasets with variable read lengths.

\subsection{System setup and configuration}\label{experiment_setup}

\subsubsection{FPGA architecture and implementation}

We have implemented \hwname{} on a Zynq UltraScale+ MPSoC platform using the Vitis HLS and Vivado Design Suite tools (v.~2022.2).
We use a ZCU104 evaluation board, which has an XCZU7EV \gls{fpga}.
The programmable part of the XCZU7EV includes over \SI{500}{\kilo\relax} logic cells, \num{1728}~DSP slices, and \SI{38}{\mega\bit} of high-bandwidth internal memory.

Additionally, the \gls{fpga} \gls{soc} includes a quad-core ARM Cortex-A53 processor,
on which we run the PYNQ image version~3.0.1 to launch and control the application.
The board has \SI{2}{\gibi\byte} of DDR4~3200 \gls{dram}, connected via a 64-bit bus to both the ARM cores and the \gls{pl}.

We implemented several optimized instances targeting different maximum sequence lengths, from \num{100}\,bp to \num{1000}\,bp, selecting the best-performing configuration for each length.
A summary of resource utilization, number of workers, and maximum frequency achieved for all implementations is presented in Table~\ref{tab:resource_utilization}.
We also adjust the query buffer to accommodate a maximum number of \num{10240} queries, except for \num{1000}\,bp maximum size due to limitation in \glspl{bram} (c.f. Section~\ref{subsec:architecture}).
Moreover, the maximum sequence length supported by our architecture is limited to \num{1000}\,bp due to Vitis HLS constraints on the maximum width for the \gls{fifo} streams.
As expected, the Worker modules dominate \gls{lut} utilization, accounting for approximately \SI{97}{\percent} of the total \glspl{lut} used for all implementations, as reported by Vitis HLS.
This design optimizes both the number of workers and the frequency of the \gls{pl}, as indicated by the number of \glspl{lut} used.

The worker critical path is dominated by the bit-vector update in the inner loop of Myers's algorithm, with the addition operation (carry propagation) setting the intrinsic limit, yielding an estimated $\mathrm{F_{max}}\!\approx\!\SI{263}{\mega\hertz}$ for an isolated worker on the ZCU104. In the full 42-worker \hwname{} configuration, the operating frequency drops to \SI{220}{\mega\hertz} due to system-level interconnect fanout and routing delays from the Split stage.

All instances achieve $\mathrm{II}=1$, with a three-cycle latency for each iteration of the inner loop of Myers's algorithm. The highest internal latency occurs in the Read module when accessing target sequences (29~cycles), which can lead to worker starvation when
\begin{equation}\label{eq:starvation}
    29 + L_{t}/4 > 2 + L_{t} - W .
\end{equation}
Here, the left-hand term represents the latency associated with reading target data in four-byte words, while the right-hand term corresponds to the number of cycles during which workers remain occupied processing the current target. This condition affects configuration targeting 100\,bp with 99~workers, where any stall in the pipeline can propagate and cause worker starvation. In contrast, for configurations targeting 200\,bp and above, the minimum waiting cycles associated with target reads do not stall the workers due to the higher sustained pipeline occupancy: $79<123$ in~(\ref{eq:starvation}).

Using the template description, Vitis HLS infers a 32\,bit AXI data width (4\,bytes), AXI burst length of~16, 16~outstanding read transactions, and 32~outstanding write transactions. We additionally specify a \SI{10}{\percent} clock uncertainty in Vitis HLS, and increased timing effort during Vivado synthesis.

\begin{table}[]
    \caption{Resource utilization of the different instances of \hwname{}}
    \centering
    \resizebox{0.48\textwidth}{!}{%
    \begin{tabular}{lrrrrr}
    \toprule
    Max seq. length (bp)     & \num{100}   & \num{200}   & \num{360}   & \num{500}   & \num{1000}  \\ 
    \midrule
    \# of workers            & \num{99}    & \num{77}    & \num{42}    & \num{29}    & \num{15}    \\
    Clock (\si{\mega\hertz}) & \num{248}   & \num{250}   & \num{220}   & \num{220}   & \num{122}   \\
    Buffer size (\# seq.)    & \num{10240} & \num{10240} & \num{10240} & \num{10240} & \num{1024}  \\
    \midrule
    BRAMs (\%)               & \num{19.87} & \num{37.50} & \num{65.54} & \num{91.67} & \num{18.59} \\
    LUTs  (\%)               & \num{57.66} & \num{87.44} & \num{86.96} & \num{81.37} & \num{90.59} \\
    FFs   (\%)               & \num{41.54} & \num{58.18} & \num{55.01} & \num{52.45} & \num{54.07} \\
    DSPs  (\%)               & \num{0.29}  & \num{0.29}  & \num{0.29}  & \num{0.29}  & \num{0.29}  \\ 
    \bottomrule
    \end{tabular}
    }
    \label{tab:resource_utilization}
\end{table}

\subsubsection{CPU setup and baseline reference}\label{sec:baselines}

To evaluate the performance of \hwname{}, we benchmark it against two state-of-the-art Myers-based accelerators, \mbox{\textit{SeqMatcher}~\cite{espinosa2025seqmatcher}} on CPU and \textit{Schifano}~\cite{schifano2025high} on \gls{fpga}. For \glspl{gpu}, we found no recent implementations of Myers's algorithm; thus, we include the older work in~\cite{chacon2014thread} as a reference. For completeness, and given the lack of recent \gls{gpu} implementations of Myers’s algorithm, we also benchmark two representative \gls{gpu}-based pairwise alignment frameworks based on alternative algorithms, \textit{WFA-GPU}~\cite{aguado2023wfa} and \textit{GASAL2}~\cite{ahmed2019gasal2}, to evaluate their performance on modern \glspl{gpu}. We also include the CPU WFA implementation \textit{WFA2-lib}~\cite{marco2021fast} to report WFA results on both CPU and \gls{gpu}. We configure all benchmarks to compute all-vs-all comparisons with no prefiltering.

An overview of the selected sequence alignment implementations, their target platforms, underlying algorithms, and scoring schemes is provided in
Table~\ref{tab:SoAcharacteris}.

\begin{table*}[tp!]
\caption{Overview of the sequence alignment implementations}
\label{tab:SoAcharacteris}
\resizebox{\textwidth}{!}{%
\begin{tabular}{cccccc}
\hline
\textbf{Aligner} & \textbf{Platform} & \textbf{Hardware} & \textbf{Algorithm} & \textbf{Scoring} & \textbf{Trace-back} \\ \hline
GASAL2~\cite{ahmed2019gasal2} & GPU & GTX 1080 Ti & Needleman-Wunsch and Smith-Waterman & gap-affine & Yes \\
WFA-GPU~\cite{aguado2023wfa} & GPU & NVIDIA GeForce RTX 3080 & WFA & gap-affine & Yes \\
Chacón et al. (2014)~\cite{chacon2014thread} & GPU & GTX Titan + Tesla K20c and 2090 & Myers & edit distance & No \\
WFA2-lib~\cite{marco2021fast} & CPU & Intel Xeon Gold 6448Y (Sapphire Rapids) & WFA & indel, edit, linear/affine/concave gap penalties & Yes \\
SeqMatcher~\cite{espinosa2025seqmatcher} & CPU & Intel Xeon Gold 6448Y (Sapphire Rapids) & Myers and Banded-Myers & edit distance & Yes \\
Schifano et al. (2025)~\cite{schifano2025high} & FPGA & Alveo U50 & Banded-Myers & edit distance & No \\
Our work & FPGA & Zynq UltraScale+ MPSoC (ZCU104, XCZU7EV) & Myers & edit distance & No \\ \hline
\end{tabular}%
}
\end{table*}

The evaluation of \textit{SeqMatcher} and \textit{WFA2-lib} is conducted on an Intel(R) Xeon(R) Gold 6448Y (``Sapphire Rapids'') server. The server has two sockets, each containing 32 physical cores with hyperthreading, \mbox{AVX-512} vector extensions, and a \gls{tdp} of \SI{225}{\watt}. The cores operate at a base frequency of \SI{2.10}{\giga\hertz} and a ``max turbo'' frequency of \SI{4.1}{\giga\hertz}. Additionally, the server is equipped with \SI{256}{\gibi\byte} of RAM. We compile \textit{WFA2-lib} with GCC using the \texttt{-O3} optimization flag; then, we utilize its C API, and execute it in unbanded mode for edit distance. Since it does not support execution without trace-back, we perform all experiments with trace-back enabled.
We compile \textit{SeqMatcher} using the Intel compiler (icpx) version 2023.1 with the flags \texttt{-mAVX-512ovf} and \texttt{-mfma}, and run it in unbanded mode without trace-back.

For the evaluation of \textit{GASAL2} and \textit{WFA-GPU}, we conduct our experiments on the partition `h100' of the KUMA cluster of EPFL.\footnote{https://top500.org/lists/green500/list/2025/11/} Each node is equipped with 4 NVIDIA H100 SXM5 \glspl{gpu}, each providing \SI{94}{\gibi\byte} of HBM2e and a memory bandwidth of \SI{2.4}{\tera\byte\per\second}.
Each compute node provides \SI{384}{\gibi\byte} of DDR5 host memory and two \SI{200}{\giga\bit\per\second} InfiniBand connections. We compile both \textit{GASAL2} and \textit{WFA-GPU} at \texttt{-O3} with NVCC for CUDA SM90 (sm\_90). The host code was compiled with G++/GCC (C++11). A single GPU of the node is used for the deployment.
We execute GASAL2 in unbanded local-alignment mode (Smith–Waterman) and we run all experiments without trace-back. 
We execute \textit{WFA-GPU} in unbanded mode and without trace-back.
As \textit{GASAL2} and \textit{WFA-GPU} do not support an explicit edit-distance mode, we configure the scoring with unit-cost penalties to improve comparability with Myers’s unit-cost edit distance. Specifically, for \textit{GASAL2} we use the scoring scheme \textit{match = +1, mismatch = -1, gap-open = 0, gap-extend = -1}, and for \textit{WFA-GPU}, we use \textit{mismatch = 1, gap-open = 0, gap-extend = 1}. Although this configuration approximates unit-cost penalties, it does not constitute a native edit-distance mode.
For the implementation of \textit{Chacón et al.}~\cite{chacon2014thread}, which is not publicly available, we use the performance values reported in their paper (c.f. Section~\ref{performance_measures}).

To mitigate result variability, we execute each CPU- and GPU-based implementation
repeatedly, with a minimum of two runs, until a total execution time of at least
\SI{100}{\second} is reached. Additionally, for  \textit{GASAL2}, \textit{WFA2-lib}, and
\textit{WFA-GPU}, we complement our own measurements with the performance values reported by Aguado-Puig et al.~\cite{aguado2023wfa} to avoid penalizing these software baselines due to differences in system configuration, data input and parameters selection.

For the \gls{fpga}-based accelerator from \textit{Schifano et al.}, which is also not publicly available, we use only the performance numbers reported in their paper, as described in Section~\ref{performance_measures}.

\subsection{Performance and energy-efficiency assessment}\label{performance_measures}

To accurately capture total energy usage, it is crucial to assess the power consumption of all the elements in the machine using complete datasets.
Partial readings may erroneously suggest greater efficiency or marginal improvements that do not accurately reflect the overall energy usage of the system, which can hinder effective decision-making in energy management strategies of data center workloads.

Therefore, to evaluate the different systems, we use two metrics, namely, speed, measured in giga cells per second (GCUPS), and energy consumption of the complete machine, measured in giga cells per joule (GCUPJ).

For the CPU- and GPU-based implementations that we run in our experiments, we measure execution time using the POSIX \textit{clock\_gettime()} function.
For \textit{Chacón et al.}~\cite{chacon2014thread}, we use the best GCUPS value reported directly in their paper.

For the work of \textit{Schifano et al.}~\cite{schifano2025high}, we consider the total matrix cells per comparison (referred as ``\#cell'' in their paper but as ``NC'' in our work), and the GCUPS reported.\footnote{When we attempted to derive the number of GCUPS by multiplying the reported values for the ``\#cell'' and the mega-pairs of sequences per second (``Mps''), our result did not match their published values. Therefore we directly use their published values for GCUPS.} Aggregating the GCUPS with the reported power (referred as ``Power Watt''), we calculate the energy consumption of their solutions.

For all platforms (CPU, \gls{gpu}, and \gls{fpga}), we measure execution time and energy over the complete workload. The total number of cell updates is denoted as $NCU$. Since targets and queries are drawn from the same dataset, $NCU$ is computed as
$NCU = N^2 \times \overline{L}^2$,
where $N$ is the number of sequences in the dataset and $\overline{L}$ is the average sequence length.

The achieved performance in giga cell updates per second (GCUPS) is defined as
$GCUPS = NCU / (t \times 10^9)$,
where $t$ is the measured execution time in seconds. Similarly, the energy efficiency in giga cell updates per joule (GCUPJ) is defined as
$GCUPJ = NCU / (e \times 10^9)$,
where $e$ is the measured energy consumption in joules.

We compute the peak compute-bound performance of \hwname{} as
\begin{equation}\label{eq:peak}
    \mathit{GCUPS}_{\mathit{peak}} = \frac{F \times W \times L_{\max}}{10^3},
\end{equation}
where $F$ is the accelerator frequency in \si{\mega\hertz}, $W$ is the number of parallel workers, and $L_{\max}$ is the average query length. The term $W \times L_{\max}$ represents the number of cell updates computed per clock cycle, and the division by $10^3$ converts from mega-updates per second to giga-updates per second.
As a numerical example, the \hwname{} instance targeting 360\,bp with $W=42$ workers performs $42 \times 360 = 15\,120$ cell updates per cycle. At an operating frequency of \SI{220}{\mega\hertz}, this corresponds to \num{3326.4}\,GCUPS.

To measure energy consumption, we utilize the Power Measurement Toolkit (PMT)~\cite{p:PMT2022}. PMT is a high-level library designed to measure energy consumption across various hardware architectures, including CPUs, \glspl{gpu}, and \as{fpga}. It interfaces with several APIs, such as NVML for NVIDIA \glspl{gpu}, ROCM-SMI for AMD \glspl{gpu}, RAPL for CPUs, and LIKWID for performance monitoring, as well as physical power sensors.

We distinguish two power measurement scopes: \emph{device-level}, corresponding to the accelerator or processor subsystem including memories; and \emph{system-level}, corresponding to the full machine executing the workload including a host device when applicable. Table~\ref{tab:power_scopes} summarizes the available power measurements by scope for each platform, and all numerical results by scope are reported in~\ref{sec:appendix}.
PMT samples the system's power periodically (we have used a sampling period of \SI{0.1}{\second}) to provide a reliable averaged metric on the total energy consumption without significantly affecting the energy consumption of the application being measured.

For CPU-based implementations, we specifically use RAPL (Running Average Power Limit) interfaced through the PMT library to monitor energy consumption.
RAPL monitors power consumption across multiple domains within Intel processors, including the package, cores, and ``uncore'' components. It uses model-specific registers (MSRs) to track energy usage over time, enabling real-time measurement of energy consumption.
In particular, we use the measurement scope labeled as ``psys,'' which provides an estimation of the total power consumed by the machine---this metric serves as the basis for our energy consumption assessments, as it includes both processor sockets and main memory. For CPU-only machines, the device-level corresponds directly to the platform-level, as the CPU constitutes the sole compute device. Using the infrastructure available at EcoCloud, we have measured a typical deviation of less than \SI{10}{\percent} with respect to the power values read by the rack's \gls{pdu} for the complete machine.

On \gls{gpu} platforms, PMT interfaces with NVIDIA’s NVML library to collect instantaneous device-level power measurements for the \gls{gpu} board. System-level measurements additionally account for the host CPU, while chip-level \gls{gpu} power is not directly observable with the available instrumentation.
For state-of-the-art implementations that are not publicly available, as in \textit{Chacón et al.}~\cite{chacon2014thread}, direct power measurements cannot be reproduced. In these cases, we report the \gls{tdp} as a best-effort proxy and explicitly indicate this limitation.

The ZCU104 evaluation kit operates as a standalone system and exposes three power meters through registers mapped within the hard-IP cores:
(i) a primary power rail supplying the \gls{dram} controller, \gls{dram}, hard-IP cores, and the \a{pl};
(ii) a secondary power rail supplying auxiliary utilities and peripherals, such as the external SD card used to boot the operating system; and
(iii) a sensor measuring the total board power supply.
We consider device-level power as the aggregation of the primary and secondary rail measurements, which account for the \gls{fpga} chip and \gls{dram}, while system-level power corresponds to the total board measurement.

\begin{table*}[ht]
\caption{Power measurement scopes available for each platform and the corresponding measurement sources}
\centering
\begin{tabular}{lll}
\toprule
\textbf{Platform} & \textbf{Device-level} & \textbf{System-level} \\
\midrule
FPGA (Standalone) &
Power rails direct meter &
Board power meter \\

FPGA (PCIe) &
Power meter device &
FPGA + host CPU \\

CPU &
System meter (RAPL \texttt{psys}) &
Same as device (CPU-only machine) \\

GPU &
GPU board power (NVML) &
GPU + host CPU \\
\bottomrule
\end{tabular}
\label{tab:power_scopes}
\end{table*}


\section{Results and analysis}
This section presents the performance and energy savings obtained by \hwname{} for the pairwise alignment problem.

\subsection{\hwname{}'s performance}
\label{performance_results}

\subsubsection{Impact of dataset scale and read-length on \hwname{} performance}

\begin{figure}[tp]
    \centering
 \includegraphics[width=\columnwidth]{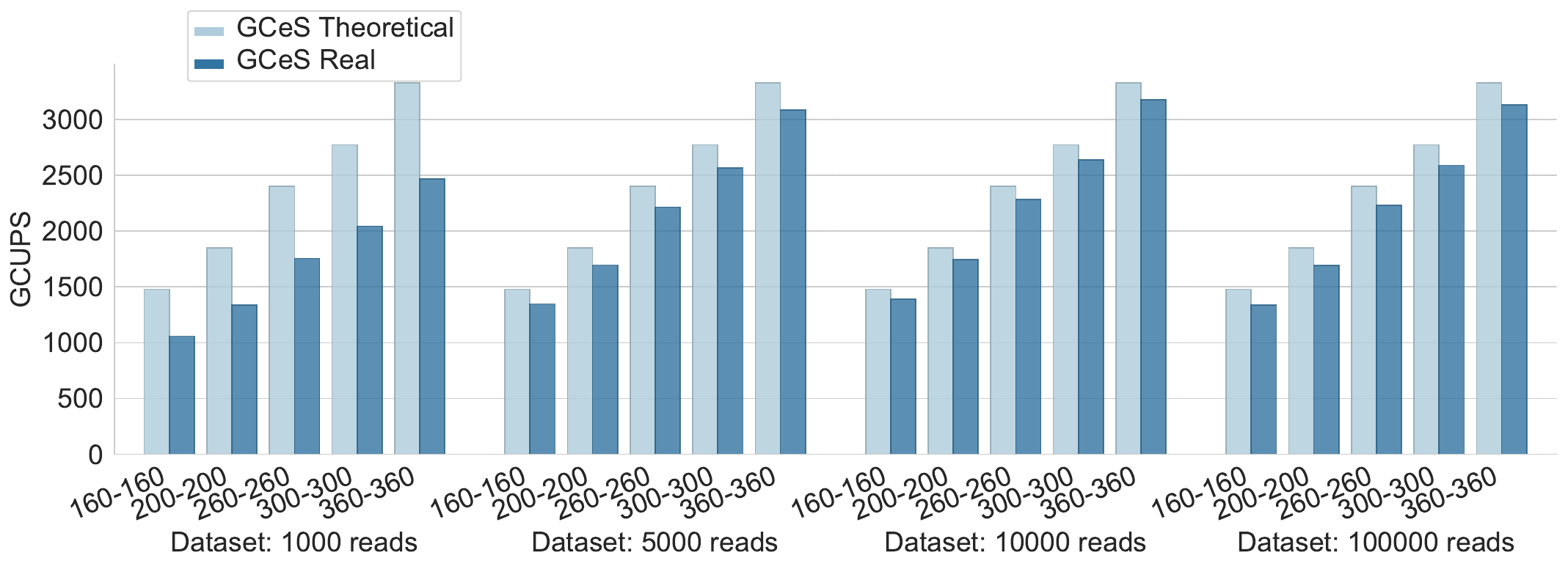}
    \caption{Comparison between theoretical and real performance, measured in giga cells per second (GCUPS), across different datasets using a fixed read length in each dataset (\textit{Group A}).} 
    \label{fig:PerfGroupA}
\end{figure}

\begin{figure}[tp]
    \centering
 \includegraphics[width=\columnwidth]{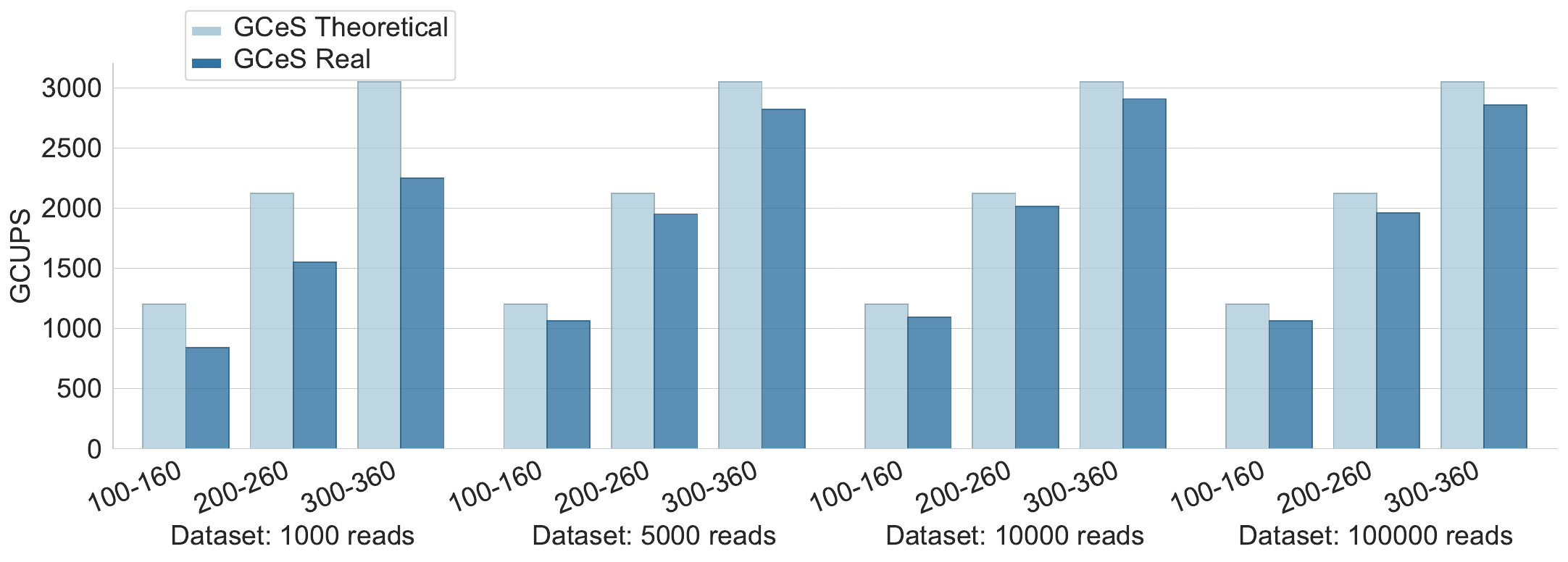}
    \caption{Comparison between theoretical and real performance, measured in giga cells per second (GCUPS), across different datasets using variable read lengths in each dataset (\textit{Group B}).}
    \label{fig:PerfGroupB}
\end{figure}

We study the effect of the number of reads and the read length on performance in terms of GCUPS. For this purpose, we report performance results for the \hwname instance that implements $L_{\max}=360$\,bp, one of the optimal operation points for the template.
Figures~\ref{fig:PerfGroupA} and~\ref{fig:PerfGroupB} present the theoretical and real numbers of GCUPS obtained for \textit{Group A} and \textit{Group B} (see Section~\ref{dataset_description}).

The theoretical performance value is calculated as the degradation of the peak performance when computing shorter sequence lengths, defined by 
\begin{equation}\label{eq:th}
    GCUPS_{th}=GCUPS_{peak} \times \frac{L_q}{L_{\max}} .
\end{equation}

When considering the impact of the number of reads on performance, in both figures \hwname{} exhibits consistent GCUPS numbers for the datasets of \num{5000}, \num{10000}, and \num{100000} reads, with a standard variation lower than \SI{2}{\percent} with respect to the mean.
Moreover, for these larger datasets, the measured values only deviate \SI{7}{\percent} with respect to the theoretical compute-bound performance, implying that with the buffering strategy, the problem is not memory-bound independently of the dataset size.
Performance degrades only in the case of \num{1000} reads, since the number of reads is \num{10}~times smaller than the buffer.
Thus, its potential cannot be fully utilized.

Regarding the effect of query size, performance increases linearly as query length increases, reaching performance levels in the order of teracell updates per second (up to \num{3175.91}~GCUPS).
This is in line with our theoretical model, where the level of parallelization achieved in the Myers's bit-vector implementation is proportional to the read length across all workers.

Thus, peak performance is achieved by datasets with fixed-length reads of \num{360} for this example case, as the workers of the instance can fully utilize their internal bit-vector parallelization.

\subsubsection{Buffer-size sensitivity analysis}

\begin{figure}[ht]
    \centering
    \includegraphics[width=0.95\linewidth,trim={1cm 0.5cm 2cm 1cm},clip]{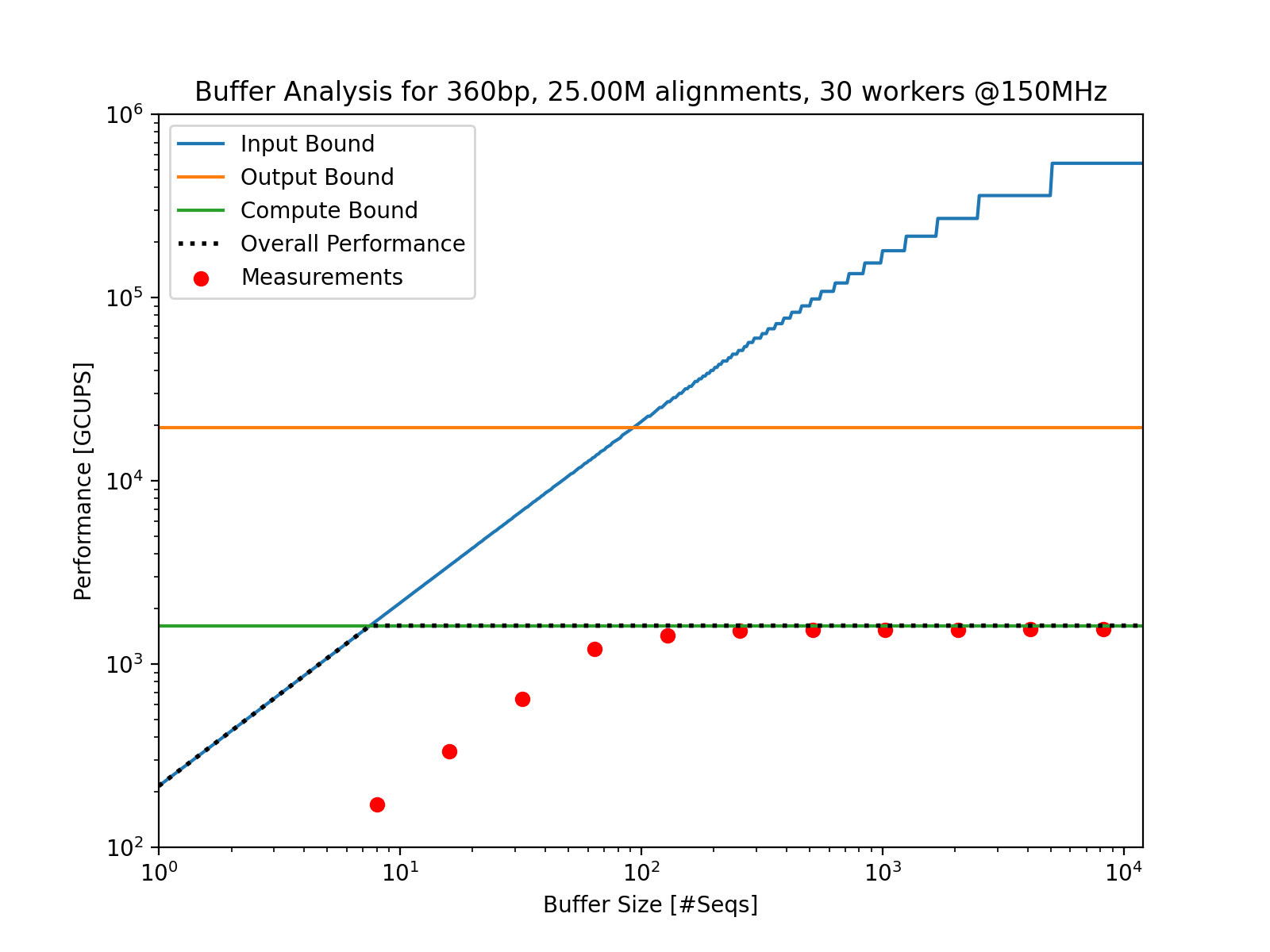}
    \caption{Buffer-size sensitivity analysis for a 360\,bp GeneTEK instance, showing the transition from memory-bound to compute-bound execution.}
    \label{fig:scalabilityB}
\end{figure}

Figure~\ref{fig:scalabilityB} reports the measured performance of a 360\,bp \hwname{} instance configured with 30~workers operating at \SI{150}{\mega\hertz} as a function of the query buffer size. The worker count and operating frequency are intentionally reduced relative to the peak-performance configuration in order to enable successful synthesis across the full range of buffer sizes under identical conditions.
For small buffers, performance is limited by input bandwidth, while beyond approximately 100~buffered queries, the design reaches the compute bound.
Although the DDR4 memory subsystem on the board provides up to \SI{19.2}{\giga\byte\per\second} of peak bandwidth, each accelerator port is configured for a maximum throughput of approximately \SI{600}{\mega\byte\per\second} per port (32-bit width at \SI{150}{\mega\hertz}). In practice, the empirical bandwidth bottleneck for small buffers is around \SI{55}{\mega\byte\per\second}. This confirms that the observed memory-bound behavior is due to accelerator-side interface and buffering constraints rather than limitations of the external memory system, and that sufficient query reuse effectively shifts execution to the compute-bound regime.

\subsubsection{Comparison of \hwname{} with other systems}\label{sec:performance}

\begin{figure}[tp]
    \centering
 \includegraphics[width=\columnwidth,trim={0 0 0cm 0},clip]{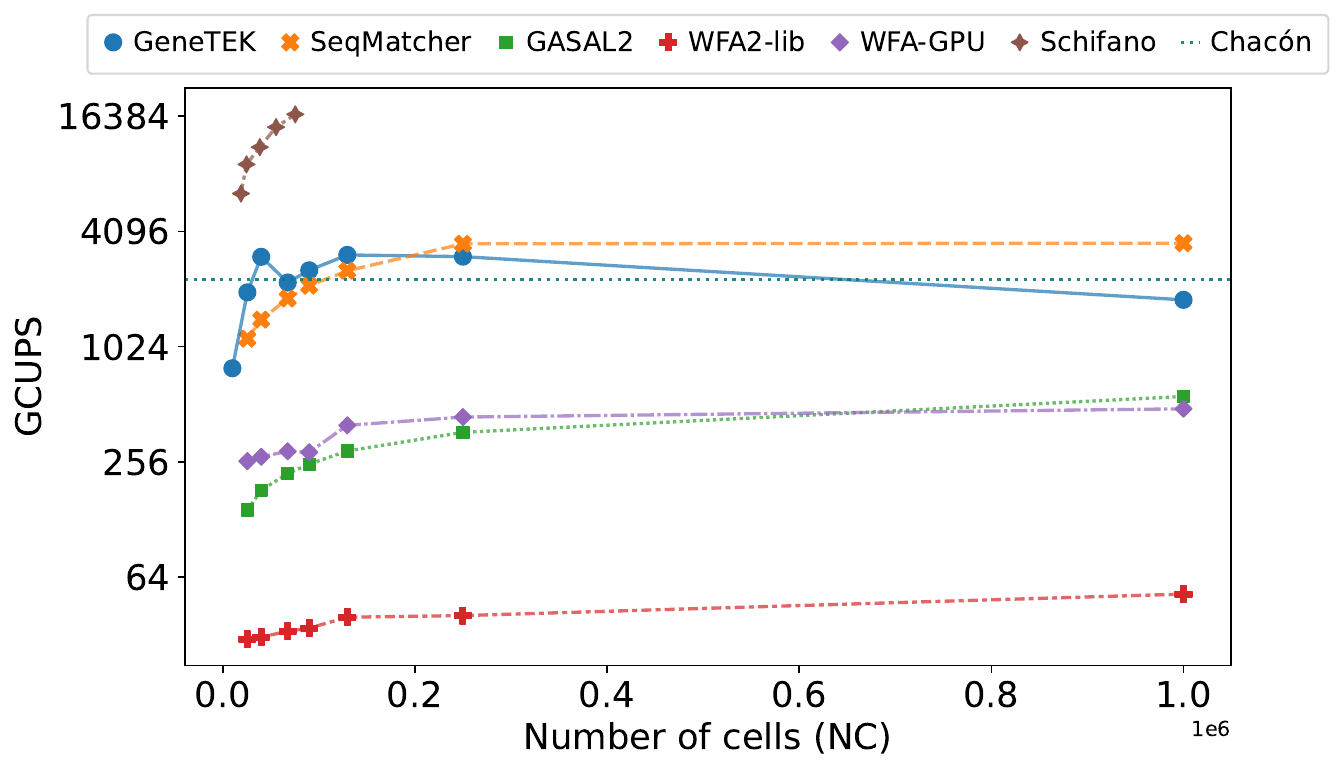}
    \caption{Performance measured in giga cells per second (GCUPS) for state-of-the-art implementations compared with \hwname{}. 
    }
    \label{fig:Fig7}
\end{figure}

Figure~\ref{fig:Fig7} compares the number of GCUPS when computing different numbers of cells in the \gls{dp} matrix (NC)---calculated as the product of the query size and the target size---for \hwname{} and for the fastest available state-of-the-art sequence alignment accelerators described in Section~\ref{experiment_setup}.
For the CPU- and \gls{gpu}-based implementations, with the exception of \textit{Chacón et al.} (see Section~\ref{performance_measures}), we report the performance measured by running each application directly on our machine. For each program, we select the thread count that yields the highest throughput. Specifically, we report results for \textit{SeqMatcher}, \textit{WFA2-lib}, and \textit{GASAL2} with 128~threads. For \textit{WFA-GPU} we use 32~CUDA threads per GPU worker. For \textit{Chacón et al.}~\cite{chacon2014thread}, since the target sequence length is not reported in the paper, we represent the highest GCUPS value. For \hwname{}, for each read size, we select the instance that achieves the highest performance and report that value. Specifically, we use the $L_{\max}=100$\,bp for reads of \num{100}\,bp; $L_{\max}=200$\,bp instance for reads of \num{160}\,bp, and \num{200}\,bp; $L_{\max}=360$\,bp for reads of \num{260}\,bp, \num{300}\,bp, and \num{360}\,bp; $L_{\max}=500$\,bp for reads of \num{500}\,bp; and $L_{\max}=1000$\,bp for reads of \num{1000}\,bp.
For both the CPU/GPU implementations and \hwname{}, the number of NC is computed using the dataset of \num{5000}~reads. Finally, for \textit{Schifano et al.}~\cite{schifano2025high}, we use their reported performance values.

We additionally report in the Figure~\ref{fig:Fig7_supp} of~\ref{sec:appendix} the number of GCUPS achieved by \textit{GASAL2}, \textit{WFA-GPU}, and \textit{WFA2-lib}, based on the throughput numbers reported by \textit{Aguado-Puig et al.}~\cite{aguado2023wfa}, as described in Section~\ref{experiment_setup}. We also report GCUPS for the approximate mode of \textit{WFA-GPU} and for the banded design of \textit{Schifano et al.}~\cite{schifano2025high} in Table~\ref{tab:banded_performance} of~\ref{sec:appendix}.

\paragraph{\hwname{} vs.\ CPU- and GPU-based accelerators.}

\hwname{} surpasses the state-of-the-art CPU and \gls{gpu} implementations of \textit{WFA2-lib}, \textit{WFA-GPU}, and \textit{GASAL2}. Furthermore, as shown in Figure~\ref{fig:Fig7_supp} of~\ref{sec:appendix}, \hwname{} also surpasses the performance of the CPU and \gls{gpu} implementations reported by \textit{Aguado-Puig et al.}~\cite{aguado2023wfa}. In addition, as reported in Table~\ref{tab:banded_performance} of~\ref{sec:appendix}, even when using the approximate mode of \textit{WFA-GPU}, \hwname{} remains faster: \hwname{} achieves a minimum throughput of \SI{789.19}{GCUPS} (see Figure~\ref{fig:Fig7}), whereas \textit{WFA-GPU} reports a maximum of \SI{480.77}{GCUPS}.
With 128~threads, \textit{SeqMatcher} outperforms \hwname{} for read lengths of 500~bp and \num{1000}~bp. In contrast, \hwname{} provides performance improvements of at least \SI{20.29}{\percent} and up to \SI{112.97}{\percent} for read lengths between 160~bp and 360~bp. Similarly, \textit{Chacón et al.}~\cite{chacon2014thread} exhibits a comparable performance considering their maximum reported throughput against the throughput of \hwname{} at each read size. However, \textit{Chacón et al.} reports much lower performance for other read sizes, down to 1~GCUPS at 100~bp~\cite{chacon2014thread}.

In general, lower performance is expected for the WFA algorithm on our dataset, as its computational cost depends on the edit distance between the aligned sequences, as discussed by \textit{Espinosa et al.}~\cite{espinosa2025seqmatcher}. Since we compute an all-vs-all workload (i.e., every read is aligned against every other read), most pairs are not candidate matches and therefore tend to have larger edit distances, which increases the alignment cost and reduces throughput. However, \textit{Aguado-Puig et al.}~\cite{aguado2023wfa} evaluate WFA on predefined sequence pairs. For simulated datasets, they generate synthetic pairs with controlled edit-error, and for real datasets they obtain target sequences by mapping the source reads against GRCh38 using Minimap2 with default parameters.
As expected, WFA2-lib achieves significantly lower throughput on our dataset than the GCUPS values reported in~\cite{marco2021fast} (see Figure~\ref{fig:Fig7_supp} in~\ref{sec:appendix}). In contrast, the GPU implementation (WFA-GPU) achieves higher GCUPS by leveraging the increased parallelism and computational throughput of a modern GPU platform (from a GeForce RTX-3080 to H100).
This makes WFA most appropriate for workloads where the query is expected to align to the target (e.g., aligning against a known genomic locus or to a predefined target reference such as an amplicon/panel), or where a prior filtering step has selected candidate targets. In contrast, the performance of \textit{SeqMatcher} and \hwname{}, both based on Myers's algorithm, remains independent of the noise level in the reads. Thus, even though WFA-GPU incorporates optimizations over WFA2-lib that reduce the performance impact of higher error rates, and even when executed on a newer GPU, its performance remains lower than \hwname{}.

Similarly, \textit{GASAL2} exhibits lower throughput on our dataset than the GCUPS values reported in~\cite{aguado2023wfa} (see Figure~\ref{fig:Fig7_supp} of~\ref{sec:appendix}). This difference could be explained given the workload organization: our experiment performs an all-vs-all comparison, whereas~\cite{aguado2023wfa} reports throughput on large, pre-formed lists of independent sequence pairs.
In all-vs-all, the $N\times N$ comparison matrix is typically processed in blocks (tiles), and each block must be packed into GPU input buffers and transferred before launching kernels. As a result, the fixed per-batch overhead (data packing, transfers, and kernel launch) is paid repeatedly, reducing end-to-end throughput. In contrast, pre-formed pair lists can be streamed in large, regular batches, which amortizes this overhead more effectively and yields higher throughput.

Although the performance of GeneTEK is not directly comparable against the performance of \textit{GASAL2} and \textit{WFA-GPU} due to the difference in scoring model, these figures are relevant in large volumes of pairwise alignments without assuming pre-matching candidates, where alignment refinement is not necessary.

\paragraph{\hwname{} vs.\ FPGA-based accelerators.}\label{systolic-vsGeneTEK}

\hwname{} uses a parallelization approach based on independent workers.
Each of the workers computes one line of the \gls{dp} matrix per cycle, thanks to the bit-vector parallelization in Myers's algorithm.
However, each worker processes only one pair of sequences per cycle.
In contrast, \textit{Schifano et al.} adopt an approach based on a fully-pipelined systolic array.
This strategy enables parallel processing of the cells in each anti-diagonal, and concurrent processing of all the anti-diagonals, which results in updating as many cells as there are in a \gls{dp} matrix in each clock cycle.
In the following paragraphs, we explore the benefits of each of these approaches, analyzing how each of them scales for increasing read lengths.

Designs based on systolic arrays reach superior performance for small read lengths, but they have strong scalability limitations for longer read lengths in unbanded computation, which restricts their domain of applicability.
The main read-length scalability problem arises from the fact that the area of the systolic array increases quadratically with the size of the input strings.
Specifically, \textit{Schifano et al.} encounter an upper limit of \num{75488}\,NC, which corresponds to very short reads (i.e., in the order of \num{250}~bases).
This significantly limits sequence mapping tasks, especially in repetitive genomes, where shorter reads increase the likelihood of incorrect mapping.
To mitigate this limitation, they employ a banded strategy to align longer reads that restricts the solution space to approximately \SI{10}{\percent} of the string length.

Nevertheless, under the reported \gls{fpga} resource constraints, the banded configurations do not yield a measurable improvement in GCUPS over the best exact approach. Thus, as shown in Table~\ref{tab:banded_performance} of \ref{sec:appendix}, \textit{Schifano et al.} report a maximum throughput of \num{15562}~GCUPS for the banded approach, whereas the unbanded approach reaches \num{16761}~GCUPS.

Specifically, they divide the input data into two groups: \textit{group1} and \textit{group2}. The former includes matrices ranging from \num{400} to \num{75488} cells. For these, the full \gls{dp} matrix is computed using a band whose width matches the sequence length. The latter consists of matrices ranging from \num{248} to \num{50888} cells. For this group, they apply a banded strategy that restricts computation to approximately \SI{10}{\percent} of the total matrix size.

Unfortunately, the banded approach introduces significant challenges in the realm of bioinformatics.
For instance, sequences that include insertions, deletions, or highly divergent regions may result in the optimal alignment straying considerably from the main diagonal.
Therefore, a pre-filtering algorithm is commonly used before band-based methods to identify candidate regions, making a fair comparison difficult in this context.

In contrast, the worker-based approach of \hwname{} improves on this read-length scalability limitation of unbanded computation by processing 13$\times$ bigger matrices, reaching \num{129600}\,NC, \emph{for a target \gls{fpga} that contains almost 4$\times$ less resources,}\footnote{The work presented in~\cite{schifano2025high} targets an Alveo U50 \gls{fpga}, which contains approximately 3.8$\times$ more \glspl{lut} and \glspl{ff} than the XCZU7EV \gls{fpga} targetted by \hwname{}. The Alveo also contains \SI{8}{\gibi\byte} of HBM accessible for the \gls{fpga} design.} 
while still implementing complete coverage of alignment. Hence, to ensure a fair comparison with the work of Schifano, we have chosen to exclude the results of \textit{group2}.

A final difference between both methods is the way in which they adapt to sequences of variable length.
Illumina sequencing typically produces fixed-length reads, although these can vary after trimming (e.g., removing adapters or low-quality bases). 
Sequences with variable length are also common in specific analysis goals such as marker gene mapping and homology comparisons.
The worker-based approach used in \hwname{} copes gracefully with these situations since the workers require a number of cycles that is proportional to the length of the target sequence.
In contrast, coping with variable lengths in systolic arrays is more complex: either the solution is propagated cycle-by-cycle until the end of the array, wasting time, or complex logic needs to be added in each stage to stop the propagation and push the local solution directly to the end of the array, where the result is output.

In summary, our analysis shows that, whereas systolic arrays achieve higher performance for shorter fixed-length sequences, worker-based approaches based on Myers's vectorization scale better to longer sequences and variable lengths.
This difference in read-length scalability is also observed in the complexity of the designs themselves, with Table~\ref{tab:designTimeAndEnergy} showing the immense difference in synthesis and implementation times for both approaches.

\subsection{Roofline analysis of the template for different max. read lengths}

\begin{figure}[ht]
    \centering
    \includegraphics[width=\linewidth,trim={1cm 0.5cm 1cm 0},clip]{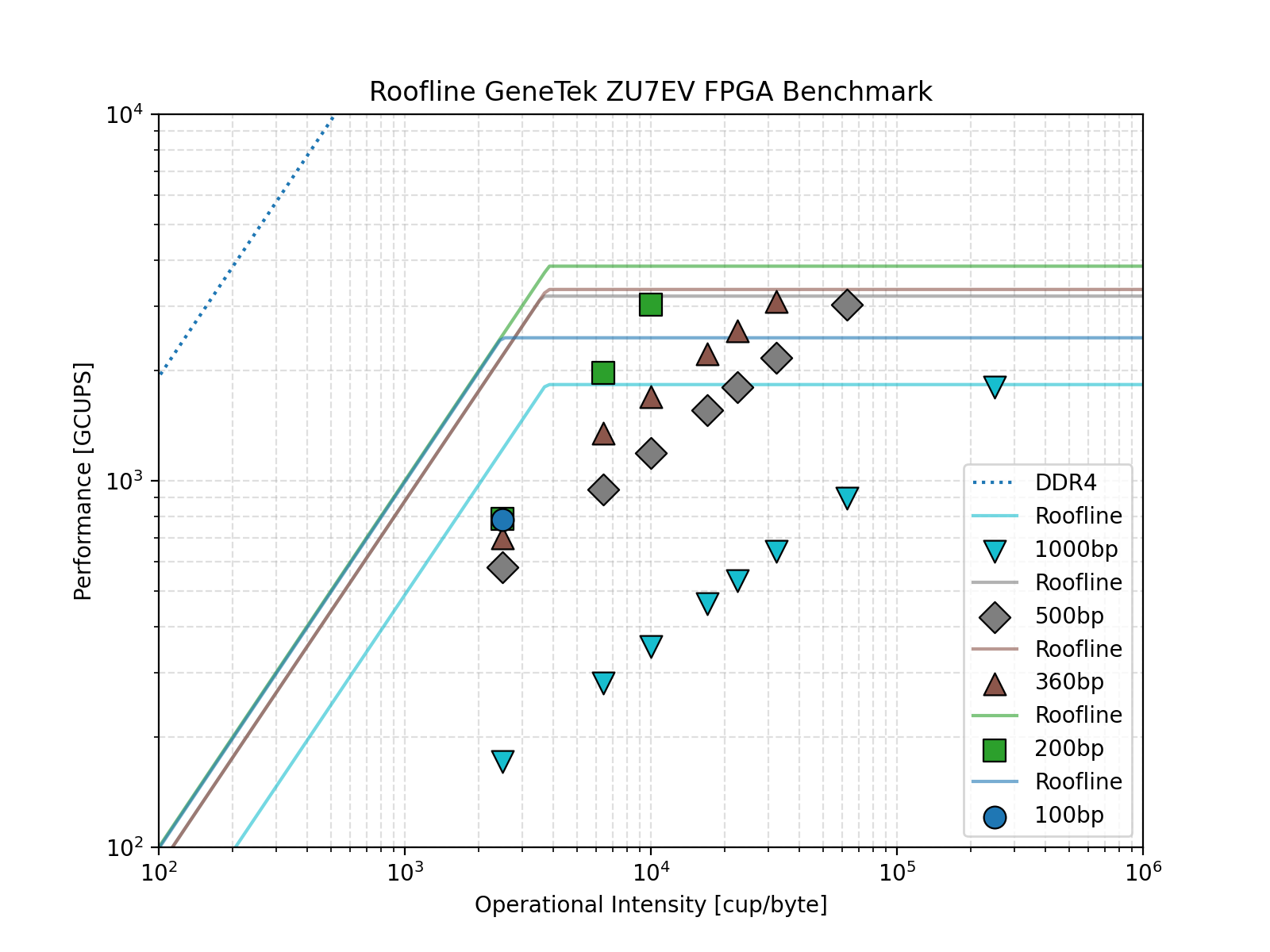}
    \caption{Roofline analysis of multiple template instances targeting different maximum sequence lengths, illustrating the transition from memory-bound to compute-bound execution as operational intensity increases and highlighting the optimal operating region of the accelerator.}
    \label{fig:roofline}
\end{figure}

Figure~\ref{fig:roofline} presents a roofline model for multiple template instances targeting different maximum sequence lengths.
The operational intensity shown corresponds to the minimum between the input (\ref{eq:oi_in}) and output (\ref{eq:oi_out}).
Instances configured between 200\,bp and \num{1000}\,bp operate close to the compute roof at sufficiently high operational intensity, confirming that these configurations are compute-bound under typical workloads.
For the instances configured to process maximum lengths of 360\,bp, 500\,bp, and \num{1000}\,bp, performance decreases approximately linearly when processing shorter sequences, as expected from reduced worker utilization.
At \num{1000}\,bp, peak performance decreases due to the limited scalability of computing resources for very large reads, although the solution remains compute-bounded.
In contrast, the 100\,bp and 200\,bp instances exhibit a pronounced deviation toward the bandwidth roof.
In these configurations, the dominant limiting factor is the output bandwidth, which is theoretically bounded by the 32-bit output port operating at \SI{248}{\mega\hertz} (approximately \SI{1}{\giga\byte\per\second}). However, only a \SI{300}{\mega\byte\per\second} output bandwidth is achieved empirically.
Overall, the measured performance points closely follow the analytical roofline model and identify an optimal operating range between 200\,bp and 500\,bp, where the accelerator avoids output bandwidth limitations while maintaining high compute utilization.

\subsection{Energy savings}

\subsubsection{Device-level}
We now analyze the energy consumption at the device-level in terms of GCUPJ of \hwname{} and the other evaluated accelerators.

\begin{figure}[tp]
    \centering
 \includegraphics[width=\columnwidth]{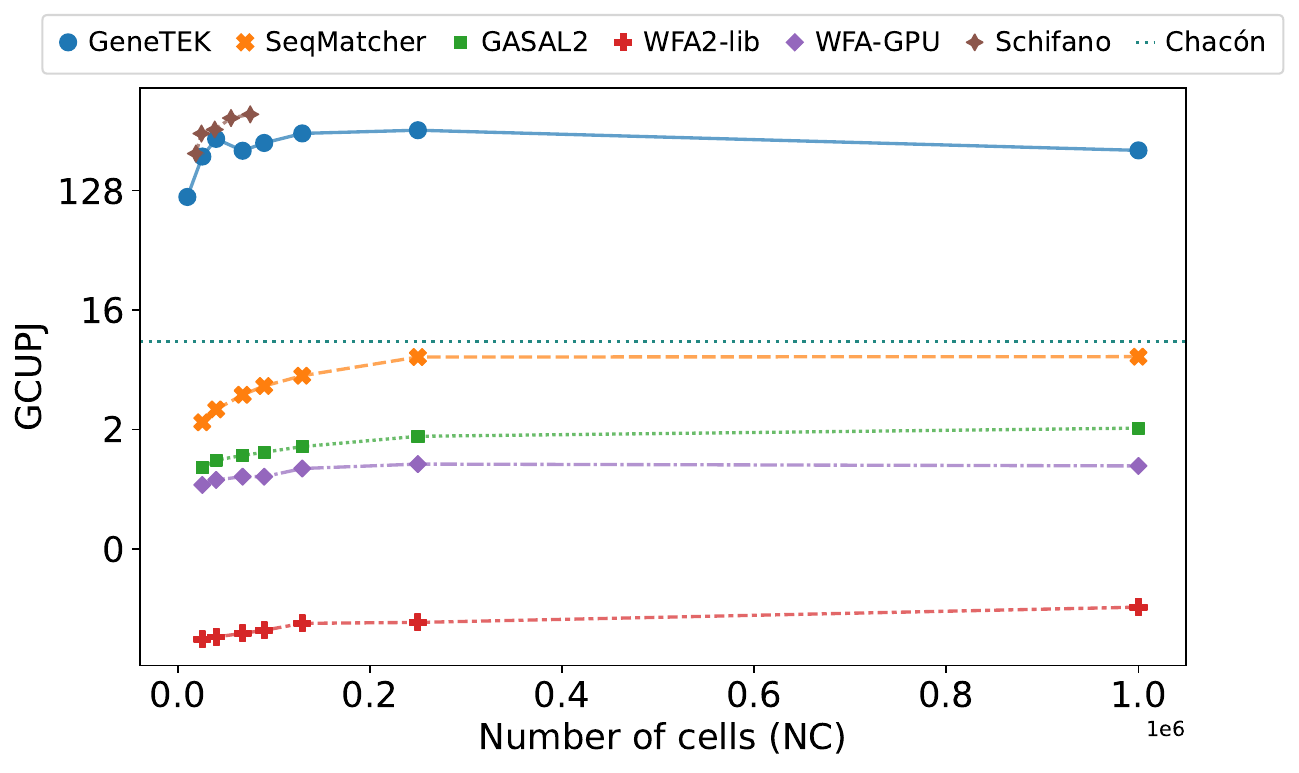}
    \caption{Energy efficiency measured in giga cells per joule (GCUPJ) with respect to the number of cells per comparison (NC) for the state-of-the-art implementations compared with \hwname{}.}
    \label{fig:Fig8}
\end{figure}

Figure~\ref{fig:Fig8} shows the number of GCUPJ for different NC values, calculated as defined in Section~\ref{performance_results}. We select the thread count that yields the highest GCUPJ value, which remains the same as in Section~\ref{sec:performance}.

Additionally, Figure~\ref{fig:Fig8_supp} of \ref{sec:appendix} reports the number of GCUPJ achieved by \textit{GASAL2}, \textit{WFA-GPU}, and \textit{WFA2-lib}, based on the throughput numbers reported by \textit{Aguado-Puig et al.}~\cite{aguado2023wfa}, as described in Section~\ref{experiment_setup}. In Table~\ref{tab:banded_performance} in \ref{sec:appendix}, we also include the corresponding GCUPJ values for the approximate mode of \textit{WFA-GPU} and for the banded design of \textit{Schifano et al.}, using the values reported in~\cite{aguado2023wfa,schifano2025high}.

In general terms, \gls{fpga}-based accelerators are significantly more energy-efficient than CPU and \gls{gpu} accelerators. 
\hwname{} delivers 36.35$\times$ to 111.41$\times$ higher energy efficiency than \textit{SeqMatcher}, up to two orders of magnitude higher energy efficiency than \textit{WFA-GPU} and \textit{GASAL2}, and more than three orders of magnitude higher energy efficiency than \textit{WFA2-lib}. In addition, Table~\ref{tab:banded_performance} in~\ref{sec:appendix} shows that this trend also holds for the approximate mode of \textit{WFA-GPU}: \hwname{} ranges between \num{114.68} and \num{366.68}\,GCUPJ (see Figure~\ref{fig:Fig8}), whereas \textit{WFA-GPU} reports at most \num{1.37}\,GCUPJ.

Interestingly, despite the high performance of \textit{SeqMatcher} in terms of GCUPS, its energy consumption is significantly higher due to the intensive use of \mbox{AVX-512} instructions. The implementation of \textit{Chacón et. al} leads to similar results: despite the higher GCUPS values achieved, its energy consumption is significantly higher.

Similarly to the GCUPS evaluation, the WFA-based baselines (\textit{WFA-CPU}/\textit{WFA-GPU}) and \textit{GASAL2} achieve lower GCUPJ on our dataset than the GCUPJ values reported in~\cite{aguado2023wfa} (plotted results in Figure~\ref{fig:Fig8_supp} from the~\ref{sec:appendix}).

Comparing the worker-based approach of \hwname{} and the systolic array approach represented by~\cite{schifano2025high}, the improvement in energy efficiency obtained by the systolic array is smaller than for performance.
Specifically, \hwname{} is only \SI{24.18}{\percent} less energy efficient than the systolic array from \textit{Schifano et al.} for the best-performing configuration in each case.
The main reason is that the average power of the \a{fpga} board used by~\cite{schifano2025high} is 2.5$\times$ higher than that of the \a{fpga} board used for \hwname{}.

Finally, consistent with the observation made for throughput in Section~\ref{systolic-vsGeneTEK}, the banded configurations in~\cite{schifano2025high} do not yield a measurable improvement in energy efficiency over their best exact design. As reported in Table~\ref{tab:banded_performance} in~\ref{sec:appendix}, \textit{Schifano et al.} achieve a maximum of \num{460.96}\,GCUPJ with the banded approach, whereas the unbanded approach reaches \num{483.72}\,GCUPJ.

\subsubsection{System-level}
An interesting consideration between the CPU/\gls{gpu} implementations and the \gls{fpga} ones is that some models of \glspl{fpga} can act as standalone systems.
Standalone \glspl{fpga} can deploy a complete application independently, using only the power required for the \gls{fpga} itself, which is typically low.

In the case of the ZCU104 development board, the additional board-level components contribute approximately 2.6\,W, corresponding to approximately 70\% higher energy consumption.

In contrast, for PCIe-mounted \gls{fpga} cards, the power of the complete host machine must be added to the power of the \gls{fpga} itself.
This consideration applies when comparing the energy consumption of \hwname{} with that of~\cite{schifano2025high}: whereas we report the energy consumed by the complete standalone board---including the accelerator, the ARM processors, the memories and all the rest of components in the standalone system---to fulfill the task, their system reports only the energy consumed by the \gls{fpga} card, without the energy consumed by the rest of the machine. 
This represents the actual energy consumption only if the system does not generate load on the server cores, and if the complete server that hosts the \gls{fpga} is concurrently being used for other task.
In this case, the system-level power is computed as

\begin{equation}\label{eq:serverOccupationFactor}
    P_{tot} = (1 - \alpha) \times P_{PC} + P_{FPGA}
\end{equation}

\noindent{}to calculate the minimum utilization factor $\alpha$ of the server that makes its overhead negligible in the computation of the \gls{fpga} system.
In particular, for a server with an idle power of \SI{100}{\watt}, the server processors must be loaded at least \SI{40}{\percent} with other tasks to compensate for the impact of the server infrastructure on the overall energy consumption.
Table~\ref{tab:energyCompleteServer} shows the total system energy consumed by \hwname{} and~\cite{schifano2025high} for varying degrees of server occupation with tasks different from the sequence matching solved in \gls{fpga}.

This overhead also impacts the reported energy consumption of GPU-based implementations. In particular, when accounting for the power of the host CPU sockets to which the GPUs are attached for the GASAL2 and WFA-GPU benchmarks, the system-level energy consumption increases by a factor of approximately 1.5$\times$ to 2.5$\times$ with respect to the device-level. This increase is primarily due to the host CPUs consuming 200--250\,W while performing minimal computational work during GPU execution (\ref{sec:appendix}).

\begin{table*}[tp]
    \caption{Complete-system energy consumption to finish a string matching task on \hwname{} and a server-based system such as~\cite{schifano2025high}. We assume an execution time of \SI{10}{\second} in \hwname{} and a performance factor of 6$\times$ for~\cite{schifano2025high}. We assume an idle-server power of \SI{100}{\watt}, an \gls{fpga} power of \SI{16}{\watt} for \hwname{}, and an \gls{fpga} power of \SI{34.65}{\watt} for~\cite{schifano2025high}. The server-based system is only advantageous if its cores can be loaded more than \SI{40}{\percent} with tasks unrelated to the sequence matching performed in the \gls{fpga}.}
    \centering
    \TableMediumFont{
    \begin{tabular}{rrr@{\hskip2cm}rrrr}
    \toprule
     \multicolumn{3}{c}{\textbf{GeneTEK}} & \multicolumn{4}{c}{\textbf{Schifano}} \\
     \textbf{Power} & \textbf{Time} & \textbf{Energy} & \textbf{Time} & \textbf{CPU utilization} & \textbf{Power} & \textbf{Energy} \\
     \textbf{(\si{\watt})} & \textbf{(\si{\second})} & \textbf{(\si{\joule})} & \textbf{(\si{\second})} & \textbf{(\si{\percent})} & \textbf{(\si{\watt})} & \textbf{(\si{\joule})} \\ \midrule
     \num{16.0} & \num{10.0} & \textbf{\num{160.0}} & \num{1.67} & \num{10} & \num{124.65} & \num{208.17} \\
     & & & & \num{40} & \num{94.65} & \textbf{\num{158.07}} \\
     & & & & \num{50} & \num{84.65} & \num{141.37} \\
     & & & & \num{100} & \num{34.65} & \num{57.87} \\
    \bottomrule
    \end{tabular}}
    \label{tab:energyCompleteServer}
\end{table*}


\section{Conclusions}

This paper proposes a SW-HW co-design approach using \gls{hls} to design \hwname{}, a new \gls{fpga}-based accelerator for pairwise sequence alignment that implements unbanded Myers's algorithm. \hwname{} achieves higher scaling of exact pairwise alignment on FPGAs than prior solutions (up to \num{1000}~bp) while maintaining energy efficiency advantages over CPU- and GPU-based platforms.
\hwname{} follows two main strategies to achieve that high performance and energy efficiency.
First, it exploits the inherent parallelism of Myers's algorithm at two levels: with a wide-bit data representation for explicit sequence vectorization, and parallelizing comparisons over multiple independent workers.
Second, it leverages the \gls{fpga}'s internal memory resources to implement data caching, removing any bottlenecks with the external memories.
The paper includes a complete roofline model analysis of the accelerator to assist in determining the performance bottleneck for each possible instantiation of the template.

The results show that the worker-based approach improves the performance and energy efficiency of state-of-the-art solutions based on CPU and \gls{gpu}, and it is competitive with previously presented \glspl{fpga}-based approaches based on systolic arrays, even when targeting smaller \gls{fpga} standalone solutions.
Additionally, this paper shows that the worker-based approach is more scalable in read lengths than the systolic array approach, both in terms of design time and complexity, and of the maximum sequence length that can be processed at run-time without resorting to banded approaches.
The presented results also show that scaling to longer sequences on FPGAs (within the evaluated length range) does not require resorting to a banded formulation.

Moreover, the experiments also show that \hwname{} outperforms state-of-the-art CPU and \gls{gpu} accelerators by achieving between \num{1092} and \num{3176}~GCUPS in datasets with \num{5000}~sequences, and an energy efficiency from \num{123} to \num{322}~GCUPJ.
Furthermore, \hwname{} achieves 13$\times$ coverage increase in NC over non-banded \a{fpga} implementations based on systolic arrays.
In conclusion, \hwname{} addresses the limitations of existing \gls{fpga}-based accelerators by offering a solution that is fast, energy-efficient, accurate, and more scalable in read lengths.


\section*{Author Contributions}

\noindent
\textbf{Conceptualization:} Elena~Espinosa, Rubén~Rodríguez~Álvarez, José~Miranda, Miguel~Peón-Quirós \\
\textbf{Methodology:} Elena~Espinosa, Rubén~Rodríguez~Álvarez, José~Miranda, Miguel~Peón-Quirós \\
\textbf{Software:} Elena~Espinosa, Rubén~Rodríguez~Álvarez, Miguel~Peón-Quirós \\
\textbf{Validation:} Elena~Espinosa, Rubén~Rodríguez~Álvarez, Miguel~Peón-Quirós \\
\textbf{Formal analysis:} Elena~Espinosa, Rubén~Rodríguez~Álvarez, Miguel~Peón-Quirós \\
\textbf{Investigation:} Elena~Espinosa, Rubén~Rodríguez~Álvarez, José~Miranda, Miguel~Peón-Quirós \\
\textbf{Visualization:} Elena~Espinosa, Rubén~Rodríguez~Álvarez \\
\textbf{Data curation:} Elena~Espinosa, Rubén~Rodríguez~Álvarez \\
\textbf{Writing – original draft:} Elena~Espinosa, Rubén~Rodríguez~Álvarez, Miguel~Peón-Quirós \\
\textbf{Writing – review \& editing:} José~Miranda, Rafael~Larrosa, Óscar~Plata, David~Atienza \\
\textbf{Supervision:} José~Miranda, Rafael~Larrosa, Miguel~Peón-Quirós, Óscar~Plata, David~Atienza \\
\textbf{Project administration:} Miguel~Peón-Quirós, Óscar~Plata, David~Atienza \\
\textbf{Resources:} Rafael~Larrosa, Óscar~Plata, David~Atienza \\
\textbf{Funding acquisition:} Miguel~Peón-Quirós, Óscar~Plata, David~Atienza \\

\section{Funding Sources}
This work has been partially supported by the Spanish MINECO PID2019-105396RB-I00, and PID2022-136575OB-I00 projects; by the EPFL Solutions 4 Sustainability program ``HeatingBits: renewable-supplied data centers integrating heating and cooling supply of local districts'', by the Swiss NSF, grant no.~200021E\_220194: ``Sustainable and Energy Aware Methods for SKA (SEAMS)'', and through a donation of material from AMD Xilinx to EPFL EcoCloud.
This article results from a research stay funded by the Spanish FPI pre-doctoral fellowship (grant PRE2020-096076) of the Ministry of Science, Innovation and Universities.

\section{Acknowledgments}
The authors also thank the computer resources, technical expertise and assistance provided by the Supercomputing and Bioinnovation Center (SCBI) of the University of Malaga.
The authors thank as well the EcoCloud center of EPFL, in particular Dr.~Xavier Ouvrard, for providing access to its infrastructure for monitoring energy consumption in servers.


\printnoidxglossary[type=\acronymtype, title={Glossary}]

\bibliography{sample}


\appendix
\renewcommand{\thesection}{Appendix \Alph{section}}
\setcounter{table}{0}
\renewcommand{\thetable}{A.\arabic{table}}

\FloatBarrier
\section{}
\label{sec:appendix}

During the development of this work, we found a particular challenge in comparing energy efficiency metrics due to the lack of information about the execution environments and tabulated results for the different benchmarks.
For this reason, as a reference for future studies, we include in Table~\ref{tab:measured} the measured results for \hwname{}, SeqMatcher~\cite{espinosa2025seqmatcher}, GASAL2~\cite{ahmed2019gasal2}, WFA-GPU~\cite{aguado2023wfa}, and WFA2-lib~\cite{marco2021fast}. Device-level power measurements are obtained following the methodology described in Section~\ref{sec:baselines}. For GPU-based implementations (WFA-GPU and GASAL), system-level energy is computed as the sum of the GPU power and the power of the two CPU sockets (``package-0'' and ``package-1''), as the \texttt{psys} domain was not available on the evaluated machines.

\begin{table}[h]
\renewcommand{\arraystretch}{0.9}
\caption{Measured time and energy results for \hwname{}, SeqMatcher~\cite{espinosa2025seqmatcher}, GASAL2~\cite{ahmed2019gasal2}, WFA-GPU~\cite{aguado2023wfa}, and WFA2~\cite{marco2021fast} for datasets of \num{5000} sequences (25\,M alignments).}
\label{tab:measured}
\scriptsize
\begin{tabular}{lrrrrr}
\toprule
Benchmark & Length & Time & Dev. Energy & Sys. Energy \\
 & (bp) & (ms) & (J) & (J) \\
\midrule
GeneTek & 200 & 331.41 & 3.17 & 5.58 \\
GeneTek & 360 & 1049.60 & 9.34 & 16.45 \\
GeneTek & 500 & 2066.61 & 17.04 & 30.58 \\
GeneTek & 1000 & 13905.60 & 96.85 & 188.38 \\
\midrule
SeqMatcher & 200 & 705.83 & 353.18 & 353.18 \\
SeqMatcher & 360 & 1270.21 & 635.81 & 635.81 \\
SeqMatcher & 500 & 1770.34 & 885.66 & 885.66 \\
SeqMatcher & 1000 & 7031.92 & 3520.25 & 3520.25 \\
\midrule
WFA2-lib & 200 & 32184.59 & 18594.25 & 18594.25 \\
WFA2-lib & 360 & 82120.82 & 47510.56 & 47510.56 \\
WFA2-lib & 500 & 155335.37 & 90186.00 & 90186.00 \\
WFA2-lib & 1000 & 479671.32 & 276506.00 & 276506.00 \\
\midrule
GASAL2 & 200 & 5498.16 & 853.92 & 2132.91 \\
GASAL2 & 360 & 11106.14 & 2186.04 & 4868.89 \\
GASAL2 & 500 & 17131.24 & 3524.06 & 7753.35 \\
GASAL2 & 1000 & 44478.21 & 12227.03 & 23721.35 \\
\midrule
WFA-GPU & 200 & 3672.70 & 1205.60 & 2191.76 \\
WFA-GPU & 360 & 8159.01 & 3202.97 & 5244.06 \\
WFA-GPU & 500 & 14246.04 & 5716.53 & 9127.11 \\
WFA-GPU & 1000 & 51612.14 & 23615.34 & 35390.85 \\
\bottomrule
\end{tabular}
\end{table}

\begin{figure}[tp]
    \centering
 \includegraphics[width=\columnwidth]{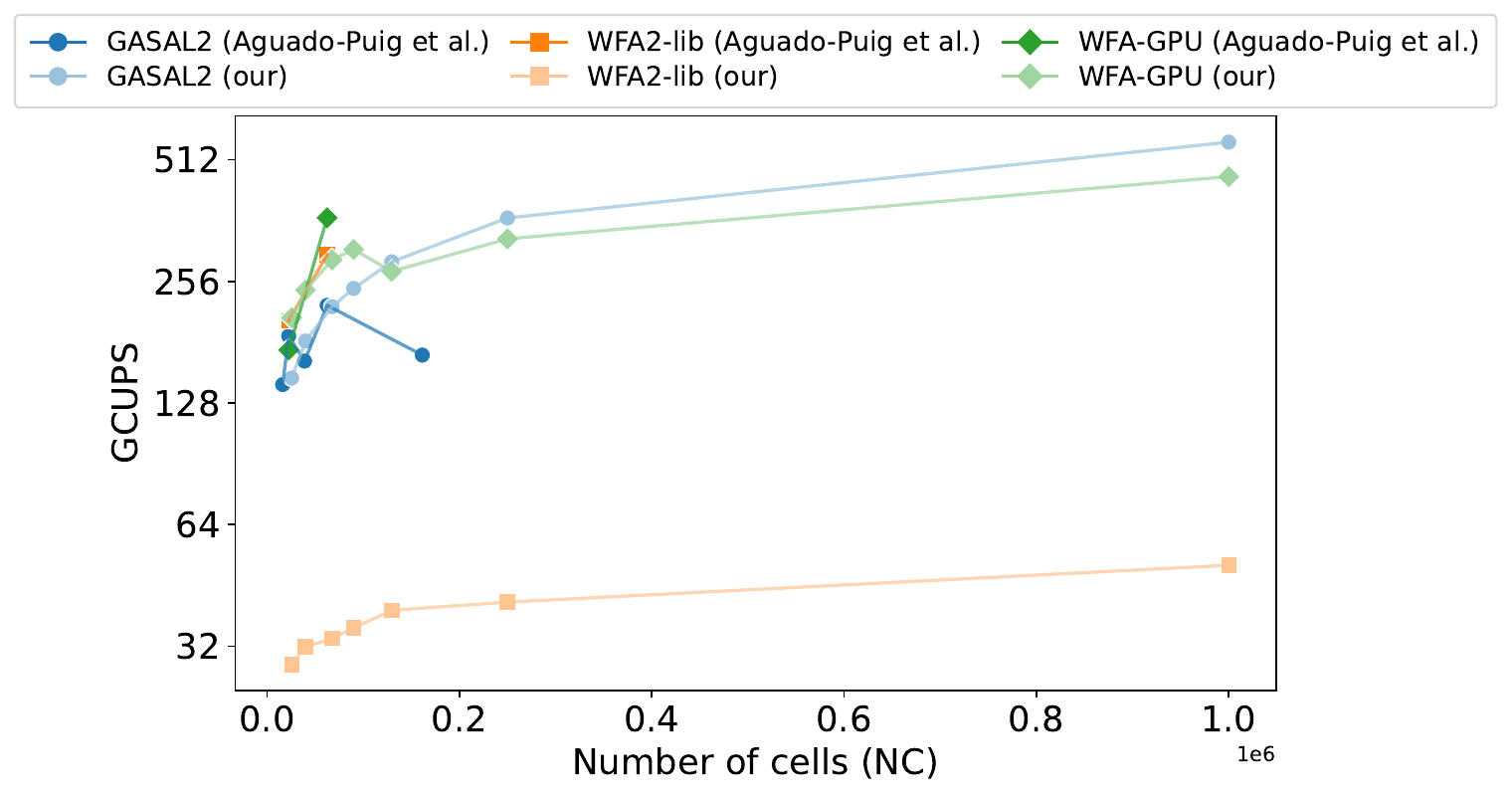}
    \caption{Performance measured in giga cells per second (GCUPS) for \textit{GASAL2}, \textit{WFA2-lib}, and \textit{WFA-GPU}, comparing our measurements with the values reported by \textit{Aguado-Puig et al.}~\cite{aguado2023wfa}. 
    }
    \label{fig:Fig7_supp}
\end{figure}

\begin{figure}[tp]
    \centering
 \includegraphics[width=\columnwidth]{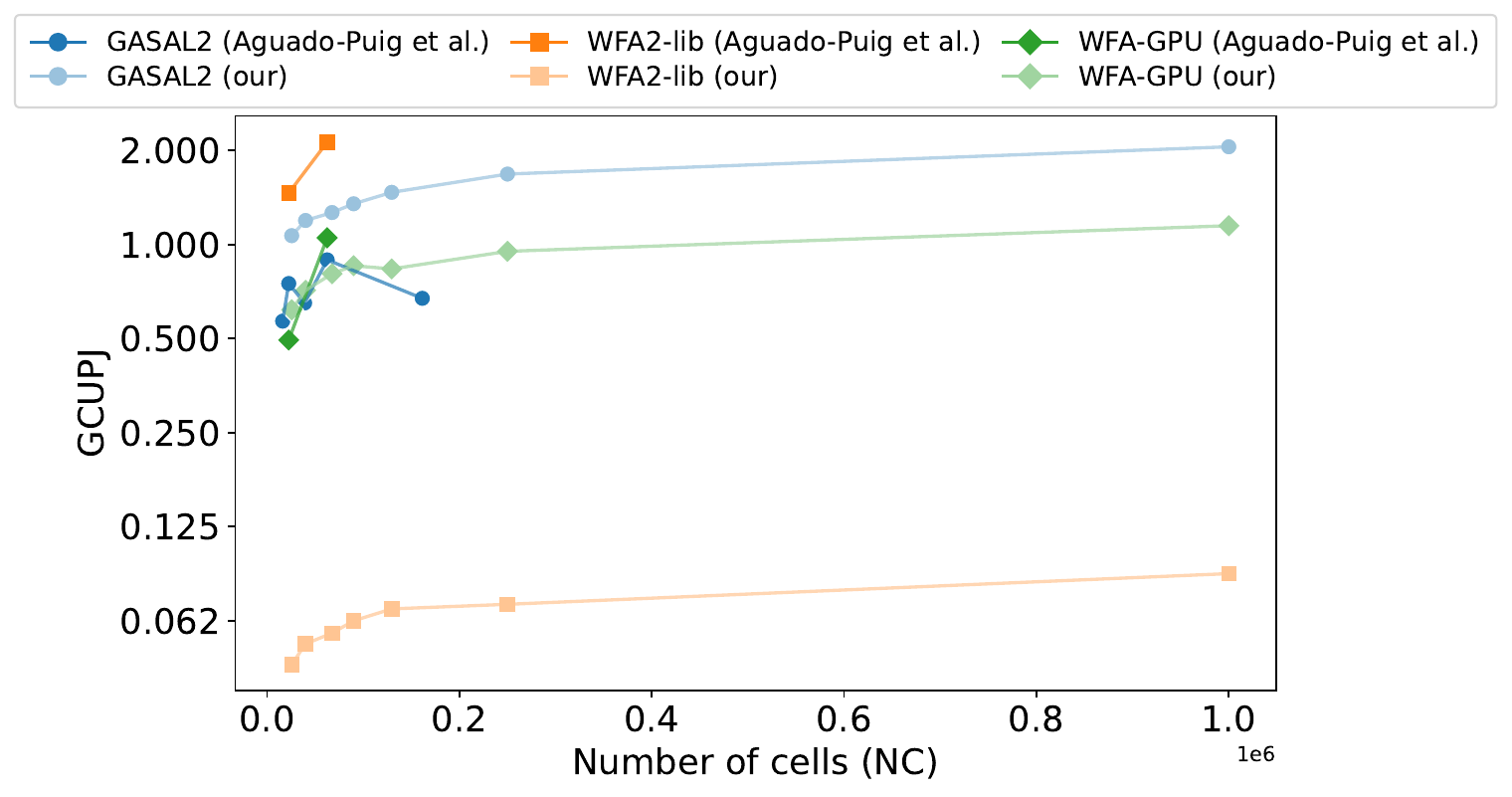}
    \caption{Performance measured in giga cells per july (GCUPJ) for \textit{GASAL2}, \textit{WFA2-lib}, and \textit{WFA-GPU}, comparing our measurements with the values reported by \textit{Aguado-Puig et al.}~\cite{aguado2023wfa}. 
    }
    \label{fig:Fig8_supp}
\end{figure}

Moreover, because the performance (GCUPS) and energy-efficiency (GCUPJ) values obtained on our dataset differ substantially from those reported by \textit{Aguado-Puig et al.}, we present both sets of results in Figures~\ref{fig:Fig7_supp} and~\ref{fig:Fig8_supp}.

\renewcommand{\thesection}{Appendix \Alph{section}}
\setcounter{table}{1}
\renewcommand{\thefigure}{A.\arabic{figure}}

Finally, Table~\ref{tab:banded_performance} reports the performance (GCUPS) and energy-efficiency (GCUPJ) results of both evaluated accelerators when executed in approximate mode.

\begin{table}[]
\caption{Performance summary of banded baselines}
\label{tab:banded_performance}
\resizebox{0.48\textwidth}{!}{%
\begin{tabular}{lrrr}
\hline
\textbf{Aligner} & \textbf{CpC} & \textbf{GCUPS} & \textbf{GCUPJ} \\ \hline
WFA-GPU~\cite{aguado2023wfa} & \num{22500} & 450.0 & 1.29 \\
WFA-GPU~\cite{aguado2023wfa} & \num{62500} & 480.8 & 1.37 \\
Schifano et al.~\cite{schifano2025high} & \num{20576} & \num{7487.0} & 271.46 \\
Schifano et al.~\cite{schifano2025high} & \num{50924} & \textsuperscript{*}\num{15562.0} & 460.96 \\
Schifano et al.~\cite{schifano2025high} & \num{51640} & \num{13503.0} & 423.16 \\ \hline
\end{tabular}%
}
\vspace{0.5em}
\footnotesize{\textsuperscript{*}Highest GCUPS reported by Schifano \textit{et al.} for their banded implementation.}
\end{table}
\end{document}